\documentclass[aip,jcp,amsmath,amssymb,reprint,]{revtex4-1}
\pdfoutput=1
\usepackage{graphicx}
\usepackage{dcolumn}
\usepackage{bm}
\usepackage{natbib}
\usepackage{gensymb}

\begin{document}

\title{Microscopic Mechanism and Kinetics of Ice Formation at Complex Interfaces: Zooming in on Kaolinite}

\author{\it Gabriele C. Sosso}
\email{g.sosso@ucl.ac.uk}
\affiliation{Thomas Young Centre, London Centre for Nanotechnology and Department of Physics and Astronomy,
University College London, Gower Street London WC1E 6BT, United Kingdom}
\author{\it Tianshu Li}
\affiliation{Department of Civil and Environmental Engineering, George Washington University, Washington, D.C. 20052, United States}
\author{\it Davide Donadio}
\affiliation{Department of Chemistry, University of California Davis, One Shields Avenue, Davis, CA 95616, USA}
\author{\it Gareth A. Tribello}
\affiliation{Atomistic Simulation Centre, Department of Physics and Astronomy, Queen's University Belfast, University Road Belfast BT7 1NN, United Kingdom}
\author{\it Angelos Michaelides}
\affiliation{Thomas Young Centre, London Centre for Nanotechnology and Department of Physics and Astronomy,
University College London, Gower Street London WC1E 6BT, United Kingdom}

\begin{abstract}

\noindent Most ice in nature forms thanks to impurities which boost the exceedingly low nucleation rate of pure
supercooled water. However, the microscopic details of ice nucleation on these substances remain largely unknown. Here,
we have unraveled the molecular mechanism and the kinetics of ice formation on kaolinite, a clay mineral playing a key
role in climate science. We find that the formation of ice at strong supercooling in the presence of this clay is twenty
orders of magnitude faster than homogeneous freezing. The critical nucleus is substantially smaller than that found for
homogeneous nucleation and, in contrast to the predictions of classical nucleation theory (CNT), it has a strong 2D
character. Nonetheless, we show that CNT describes correctly the formation of ice at this complex interface.  Kaolinite
also promotes the exclusive nucleation of hexagonal ice, as opposed to homogeneous freezing where a mixture of cubic and
hexagonal polytypes is observed.

\end{abstract}

\maketitle

The formation of ice is at the heart of intracellular freezing~\cite{tam_solution_2009}, stratospheric ozone
chemistry~\cite{bolton_ice-catalyzed_2001},  cloud
dynamics~\cite{tang_interactions_2016}, rock weathering~\cite{bartels-rausch_ice_2012} and hydrate
formation~\cite{pirzadeh_molecular_2013}.  As ice nucleation within pure supercooled liquid water is amazingly rare in
nature, most of the ice on Earth forms heterogeneously, in the presence of foreign particles which boost the ice
nucleation rate~\cite{murray_ice_2012}.  These substances, which can be as diverse as
soot~\cite{lupi_heterogeneous_2014}, bacterial fragments~\cite{osullivan_relevance_2015} or mineral
dust~\cite{eastwood_ice_2008}, lower the free energy barrier for nucleation and make ice formation possible
even at a few degrees of supercooling. 
However, the microscopic details of heterogeneous ice nucleation are still poorly understood. State-of-the-art
experimental techniques can establish whether a certain material is efficient in promoting heterogeneous ice nucleation,
but it is very challenging to achieve the temporal and spatial resolution required to characterize the process at
the molecular level. On the other hand, spontaneous fluctuations that produce nuclei of critical size are \textit{rare
events}. They thus happen on timescales (seconds) that are far beyond the reach of classical molecular dynamics
simulations.  This is why, to our knowledge, quantitative simulations of heterogeneous ice nucleation have been
successful only when using the coarse grained mW model for
water~\cite{lupi_heterogeneous_2014,fitzner_many_2015,cabriolu_ice_2015}.
Such simulations have gone a long way towards improving our fundamental understanding of heterogeneous ice nucleation, but coarse grained 
models are not appropriate for many of the more complex and interesting ice nucleating substrates.

A representative example is the formation of ice on clay minerals - a phenomenon critical to cloud formation and
dynamics~\cite{zimmermann_ice_2008,eastwood_ice_2008}.
For instance, the heterogeneous ice nucleation probability in the presence of kaolinite, a clay mineral well
studied by both
experiments~\cite{murray_heterogeneous_2011,murray_ice_2012,tobo_impacts_2012,welti_exploring_2013,wex_kaolinite_2014}
and simulations~\cite{hu_kaolinite_2010,tunega_ab_2004,cox_microscopic_2014,zielke_simulations_2016}, seems to be
related to its surface area~\cite{murray_heterogeneous_2011}, but 
how exactly this material facilitates the formation of the ice nuclei is largely
unestablished.
Classical molecular dynamics simulations have recently succeeded in simulating ice nucleation on kaolinite~\cite{cox_microscopic_2014,zielke_simulations_2016}. 
However, finite size
effects~\cite{cox_microscopic_2014} and rigid models of the surface~\cite{zielke_simulations_2016} prevented the
extraction of quantitative results.  In fact, it is exceedingly challenging to compute via atomistic simulations ice
nucleation rates, which have been inferred (for homogeneous freezing only) along a wide range of temperatures~\cite{sanz_homogeneous_2013} 
and recently computed directly at strong supercooling ($\Delta T$=42 K) for
the fully atomistic TIP4P/Ice model of water~\cite{haji-akbari_direct_2015}.

In this work, we have computed the rate and unraveled the mechanism at the all-atom level of the heterogeneous
nucleation of ice. We have considered the hydroxylated (001) surface of kaolinite as a prototypical material
capable of promoting ice formation. We quantify the efficiency of kaolinite in boosting ice nucleation and
find that this mineral alters the ice polytype that would form homogeneously at the same conditions.
We also observe that ice nuclei grow in a non-spherical fashion, in contrast with the predictions of Classical Nucleation Theory (CNT) which
nonetheless we demonstrate is reliable in describing quantitatively the heterogeneous nucleation process. 

Kaolinite (Al$_2$Si$_2$O$_5$(OH)$_4$) is a layered aluminosilicate, in which each layer contains a tetrahedral silica
sheet and an octahedral alumina sheet -- in turn terminated with hydroxyl groups.  Facile cleavage along the (100) basal
plane parallel to the layers results in surfaces exposing either the silicate terminated face or the hydroxyl-terminated
one. The latter is believed to be the most effective in promoting ice nucleation, as the hydroxyl groups form a
hexagonal arrangement that possibly templates ice formation~\cite{pruppacher_microphysics_1997,cox_microscopic_2014}.
Here we considered a single slab of kaolinite cleaved along the
(100) plane so that it exposes the hydroxyl-terminated surface, while water molecules have been represented by the fully
atomistic TIP4P/Ice model~\cite{abascal_potential_2005}.  Further details about the structure of the water-kaolinite interface and the computational setup can be found in
Refs.~\citenum{hu_water_2008,cox_microscopic_2014} and in the Supporting Information (SI). 

The heterogeneous ice nucleation rate was obtained using the Forward Flux Sampling (FFS) technique~\cite{allen_forward_2009},
which has been successfully applied for homogeneous water freezing~\cite{li_homogeneous_2011,li_ice_2013} and for
diverse nucleation
scenarios~\cite{valeriani_rate_2005,wang_homogeneous_2009,bi_probing_2014,gianetti_computational_2016}.  Within this
approach, the path from liquid water to crystalline ice is described by an order parameter $\lambda$.  A set of
discrete interfaces characterized by an increasing value of $\lambda$, is identified along this
order parameter. Here, we have chosen $\lambda$ as the number of water molecules in the largest ice-like cluster within
the whole system plus its first coordination shell (see SI). The natural fluctuations of the system at each interface, sampled by
a collection of unbiased molecular dynamics simulations, are then exploited and the nucleation rate $J$ is calculated
using:

\begin{equation}
J={\Phi}_{\lambda_0}\prod_{i=1}^{N_{\lambda}} P (\lambda_i|\lambda_{i-1})
\end{equation}

\noindent where ${\Phi}_{\lambda_0}$ is the rate at which the system reaches the first interface $\lambda_0$.
The total probability $P (\lambda|\lambda_{0})$ for a trajectory starting from $\lambda_0$ to reach the ice basin is decomposed into the product
of the crossing probabilities $P (\lambda_i|\lambda_{i-1})$.  The details of the algorithm are described in the SI.

\begin{figure}[t!]
\centerline{\includegraphics[width=7cm]{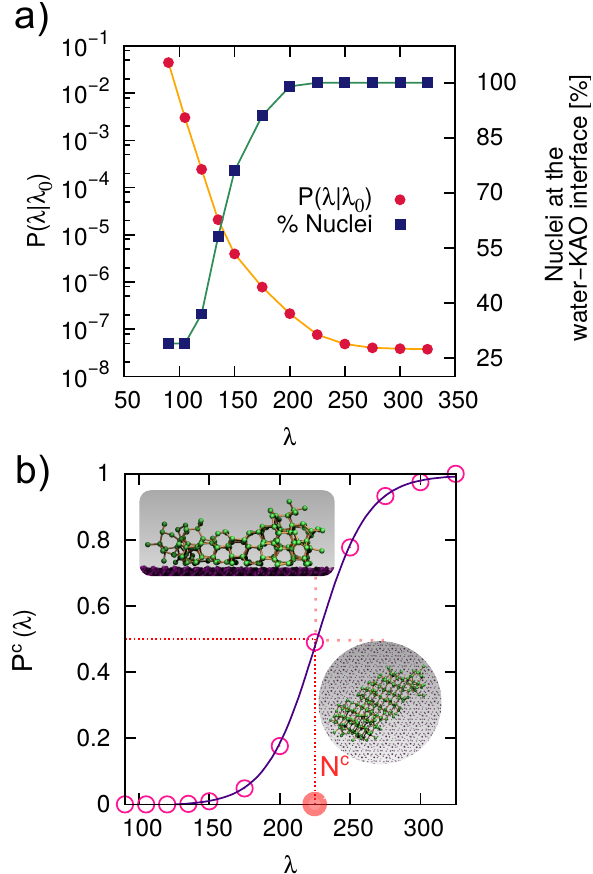}}
\caption{a) Calculated growth probability $P(\lambda|\lambda_0)$ and fraction of ice nuclei sitting on top of the
kaolinite (001) hydroxylated surface as a function of $\lambda$.  b) Committor probability $P^C(\lambda)$ as a function
of $\lambda$.  The value of $P^C(\lambda)$=0.5, corresponds to the critical nucleus size $N^C$=225.  A typical ice
nucleus of critical size is shown in the insets.} \label{FIG_1} \end{figure}

In order to compare our results with the homogeneous data from Ref.~\citenum{haji-akbari_direct_2015}, we have
performed FFS simulations at the same temperature T=230 K, corresponding for the TIP4P/Ice model to $\Delta T$=42 K.  The calculated growth
probability $P (\lambda|\lambda_{0})$ as a function of $\lambda$ is reported in Fig.~\ref{FIG_1}a.
In contrast with the transition probability for homogeneous nucleation reported in
Ref.~\citenum{haji-akbari_direct_2015}, we do not observe any \textit{inflection region}, i.e. a regime for which the
$P (\lambda|\lambda_{0})$ decreases sharply ($P(\lambda_i|\lambda_{i-1}) < P(\lambda_j|\lambda_{j-1})$ for $i<j$). 
This inflection is because in the early stages of homogeneous nucleation the largest nuclei are mostly made of hexagonal ice ($I_h$),
which leads to rather aspherical nuclei that are very unlikely to survive and reach the later parts of the nucleation
pathway. 
Within the inflection region the nuclei contain a substantial fraction of cubic ice ($I_c$).
It seems that in forming this polytype the nuclei are able to adopt a more spherical shape and that this is essential
for ultimately growing toward the critical nucleus size. In contrast, within this heterogeneous case, the presence of the surface allows 
this process of forming spherical
$I_c$-rich crystallites to be bypassed. Here, nucleation proceeds exclusively heterogeneously at the kaolinite-water
interface.  During the early stages of the process the fraction of ice nuclei on the surface (as defined in the SI) is only around 25\%,
as shown in Fig.~\ref{FIG_1}a, since at this strong $\Delta T$ natural fluctuations toward the ice phase are abundant and
homogeneously distributed throughout the liquid.  However, as nucleation proceeds the nuclei within the bulk of
the liquid slab become less favorable, until only nuclei at the water-kaolinite interface survive. From
this evidence alone one can conclude that at this temperature kaolinite substantially promotes the formation of ice via
heterogeneous nucleation. 

\begin{figure}[t!]
\centerline{\includegraphics[width=7cm]{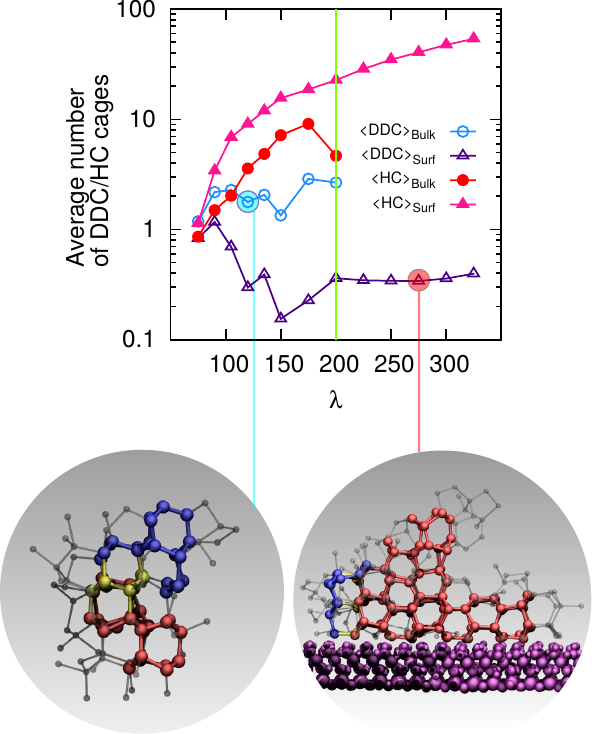}}
\caption{Average number of Double-Diamond Cages $\langle DDC \rangle_{Bulk}$
and Hexagonal Cages $\langle HC \rangle_{Bulk}$ within the largest ice nuclei (identified according to the order parameter $\lambda$) 
in the bulk of the liquid slab only as a function
of $\lambda$ (nuclei in the bulk disappear beyond the value of $\lambda$ marked by the vertical green line).  Averages
for the largest ice nuclei sitting on top of the kaolinite (001) hydroxylated surface ($\langle DDC \rangle_{Surf}$ and
$\langle HC \rangle_{Surf}$) are also reported.  The insets depict DDC and HC within an ice nucleus in the bulk at the
early stages of nucleation (left) and a post-critical ice nucleus at the water-clay surface (right).  
Oxygen atoms belonging to the largest ice nucleus (hydrogens not shown) are depicted in
blue (DDC), red (HC) and yellow (both DDC and HC). Atoms belonging to the largest ice nucleus but not involved in any DDC or HC
are shown in gray.} \label{FIG_2}
\end{figure}

Our FFS simulation results in a heterogeneous ice nucleation rate of $J_{Hetero}$=$10^{26 \pm 2}$ s$^{-1}$m$^{-3}$, which
can be compared with the homogeneous nucleation rate of $J_{Homo}$=$10^{5.9299\pm 0.6538}$ s$^{-1}$m$^{-3}$
reported in Ref.~\citenum{haji-akbari_direct_2015}. The hydroxylated (001) surface of kaolinite thus enhances the
homogeneous ice nucleation rate by about twenty orders of magnitude at $\Delta T$=42 K. This spectacular boost is
similar to that reported for simulations of heterogeneous ice nucleation on graphitic surfaces~\cite{cabriolu_ice_2015}
and on Lennard-Jones crystals~\cite{fitzner_many_2015} at similar $\Delta T$ using the coarse grained mW model.  

An estimate of the critical nucleus size $N^C$ can be obtained directly from the crossing probabilities assuming that
$\lambda$ is a good reaction coordinate for the nucleation process~\cite{haji-akbari_direct_2015}. In this scenario
$N^C$ is the value for which the committor probability $P^C(\lambda)$ for the nuclei to proceed towards the ice phase
instead of shrinking into the liquid is equal to 0.5. As shown in Fig.~\ref{FIG_1}b, $P^C(\lambda)$=0.5 corresponds in
our case to a critical nucleus of 225 $\pm$ 25 water molecules.  The estimate of the homogeneous critical nucleus size,
obtained by means of the same approximate approach employed here, is $N^C$=500 $\pm$ 30 water molecules (as obtained by
using the definition of $\lambda$ employed in this work, see SI), more than two times larger than our estimate for the
heterogeneous case. 

At this supercooling, homogeneous water nucleates into stacking disordered ice (a mixture of $I_h$ and
$I_c$)~\cite{moore_is_2011,hansen_formation_2008,malkin_stacking_2014}.  However, the presence of the clay leads to a
very different outcome. To analyze the competition between $I_h$ and $I_c$ we have adopted the topological criterion
introduced in Ref.~\citenum{haji-akbari_direct_2015} (see SI), pinpointing the building blocks of $I_c$ (Double-Diamond
Cages, DDC) and $I_h$ (Hexagonal Cages, HC) within the largest ice nuclei. The results are summarized in Fig.~\ref{FIG_2}: for ice nuclei in the bulk,
a slightly larger fraction of HC with respect to DDC develops until they disappear because of the dominance of the much
more favorable nuclei at the surface. In contrast, nuclei at the surface contain a large fraction of HC from the
earliest stages of the nucleation, and they exclusively expose the prism face of $I_h$ to the hexagonal arrangement of
hydroxyl groups of the clay. This is consistent with what has been suggested previously by classical MD
simulations~\cite{cox_microscopic_2014,zielke_simulations_2016}, and demonstrates that at this supercooling
heterogeneous nucleation takes place solely via the hexagonal ice polytype, in contrast with homogeneous nucleation.
Experimental evidence~\cite{malkin_stacking_2014} suggests that stacking disordered ice on kaolinite is likely to appear
after the nucleation process due to the kinetics of crystal growth and the presence of surfaces other than the
hydroxylated (001).

\begin{figure}[t!]
\centerline{\includegraphics[width=7cm]{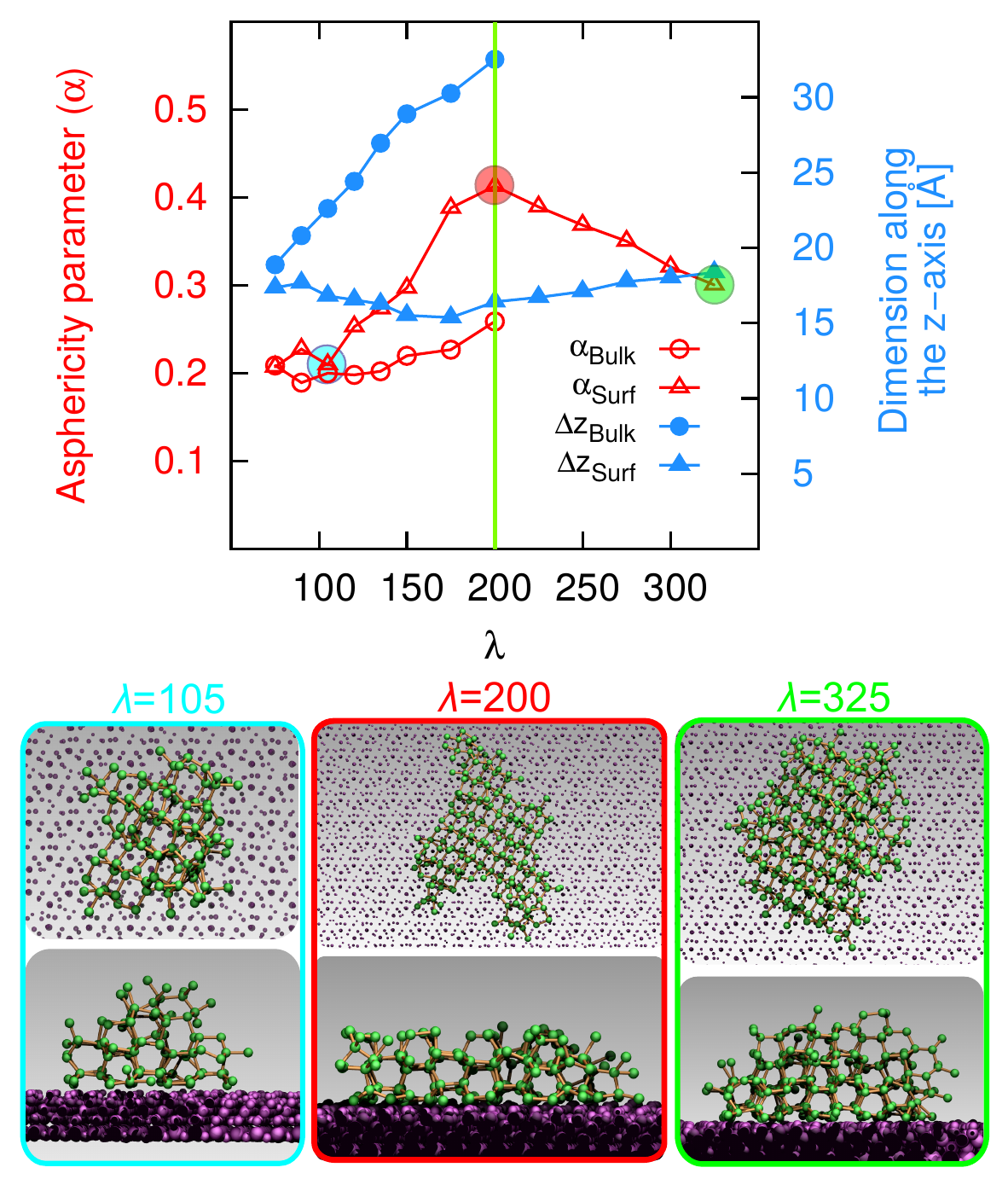}}
\caption{Asphericity parameter $\alpha$ and spatial extent of the ice nuclei along the direction normal to the clay slab
$\Delta z$ as a function of $\lambda$ for ice nuclei in the bulk ($\alpha_{Bulk}$ and $\Delta z_{Bulk}$).  Nuclei in the
bulk disappear beyond the value of $\lambda$ marked by the vertical green line.  Averages within the ice nuclei sitting on
top of the kaolinite (001) hydroxylated surface ($\alpha_{Surf}$ and $\Delta z_{Surf}$) are also reported.  The insets
correspond to typical ice nuclei containing about 105, 200 and 325 (from left to right) water molecules.}  \label{FIG_3}
\end{figure}

In the homogeneous case, critical nuclei tend to be rather spherical even at this strong
supercooling~\cite{haji-akbari_direct_2015}. However, we see a very different behavior here.  This is illustrated in
Fig.~\ref{FIG_3}, where we show as a function of $\lambda$ the asphericity parameter $\alpha$ (which is equal to zero
for spherical objects and one for an infinitely elongated rod), for nuclei in the bulk and at the surface.  Note that
heterogeneous CNT predicts (on flat surfaces) critical nuclei in the form of spherical caps, the exact shape of which is
dictated by the contact angle $\theta_{Ice,Surf}$ between the nuclei and the surface~\cite{cabriolu_ice_2015}. For
instance, $\alpha$=0.094 for a pristine hemispherical cap, corresponding to $\theta_{Ice,Surf}$=90\degree.  Also
reported in Fig.~\ref{FIG_3} is the spatial extent $\Delta z$ of the nuclei along the direction normal to the slab (the
exact definitions of $\alpha$ and $\Delta z$ are provided in the SI).  Nuclei within the bulk tend to be rather
spherical.  A small increase in the asphericity is observed right before these nuclei disappear and are replaced with
nuclei at the surface. This regime, in which the nuclei in the bulk grow substantially and become less spherical,
possibly corresponds to the onset of the inflection region observed within the homogeneous case.  However, here
nucleation is dominated by the surface. While nuclei at the surface are initially quite similar to spherical caps, they tend to grow by
expanding at the water-kaolinite interface because of the favorable templating effect of the hydroxyl groups, which
favors the formation of the prism face of $I_h$~\cite{cox_microscopic_2014}. This can clearly be seen by looking at the
substantial increase in $\alpha$ for the nuclei at the surface, which is accompanied by a slight drop in $\Delta z$
corresponding to an expansion of the nuclei in two dimensions.  Once the nuclei have overcome the critical nucleus size,
they tend to return to a more isotropic and compact form, while accumulating new ice layers along the normal to the
surface. We note that due to the strong two-dimensional nature of the critical ice nuclei, special care has to be taken to avoid 
finite size effects. We have therefore used a simulation box with lateral dimensions of the order of 
60 \AA, which is large enough to prevent interactions between the ice nuclei and their periodic images (as discussed in the SI).

The fact that the system reaches the critical nucleus size by expanding chiefly in two dimensions is in sharp contrast
with the heterogeneous nucleation picture predicated by CNT. Hence the question arises: \textit{Is CNT able to describe
heterogeneous ice nucleation on a complex substrate at this strong supercooling?} Strikingly, the answer is yes. In
order to show this, we compare the shape factor for heterogeneous nucleation $\mathcal{F}_S=\Delta G^C_{Hetero} / \Delta
G^C_{Homo}$, customarily used in CNT~\cite{sear_nucleation:_2007} to quantify the net effect of the surface on the free
energy barrier for nucleation $\Delta G^C$, with the volumetric factor $\mathcal{F}_V=N^C_{Hetero} / N^C_{Homo}$.
Details of this comparison are included in the SI. 
Note that while different, equally valid ways of defining an ice-like cluster can lead to different values of $N^C$, there is no
ambiguity in the estimate of $\mathcal{F}_S$ and $\mathcal{F}_V$ as long as the same order parameter is used to
define both $N^C_{Homo}$ and $N^C_{Hetero}$.
Thus, we obtain $\mathcal{F}_S=0.4\pm 0.1$, in very good agreement with
$\mathcal{F}_V=0.45 \pm 0.08$.  Heterogeneous CNT has already proven to be reliable in describing the crystallization of
ice on graphitic surfaces~\cite{cabriolu_ice_2015}, a scenario very different from ice formation on kaolinite.  In fact,
while the size of the critical clusters reported in Ref.~\citenum{cabriolu_ice_2015} is similar to what has been
obtained here (few hundreds of water molecules), critical ice nuclei of mW water on graphitic surfaces are shaped as
spherical caps, in line with CNT assumptions. This is due to the fairly weak interaction between water and carbonaceous
surfaces~\cite{lupi_heterogeneous_2014}, which results in a weak wetting of the ice phase on the substrate. In contrast,
our results show that ice nuclei on kaolinite tend to wet substantially the substrate, leading to shapes very different
from spherical caps.  In this regime, where the nuclei are small and the ice-kaolinite contact angle
$\theta_{Ice,Surf}$ is also small, line tension at the water-ice-kaolinite interface could introduce a mismatch between
$\mathcal{F}_S$ and $\mathcal{F}_V$ (see e.g.  Refs.~\citenum{auer_line_2003,winter_monte_2009}).  However, this is not
the case, as CNT holds quantitatively for the formation of ice on kaolinite even at the strong supercooling probed in
this work.

The value of $J_{Homo}$ reported in Ref.~\citenum{haji-akbari_direct_2015} is about eleven orders of magnitude smaller
than the experimental value extrapolated from Ref.~\citenum{sellberg_ultrafast_2014}. In addition, at the strong
supercooling of $\Delta T$=42 K no direct measures of $J_{Hetero}$ exist for kaolinite (nor indeed for the homogeneous
case), as pure water freezes homogeneously at $T<\Delta T \sim 38 K $.  Consequently, extrapolations are necessary, leading to experimental
uncertainties as large as six orders of magnitude~\cite{sosso_cnm}. Nonetheless, our $\mathcal{F}_S$
quantifies the $relative$ ice nucleation ability of kaolinite with respect to the homogeneous case, which can thus
be compared with experimental values. Estimates of $\mathcal{F}_S$ from measurements of ice formation on kaolinite
particles can vary from 0.23 to 0.69 according to the interpretation of the experimental data~\cite{ickes_classical_2016}, and
the seminal work of Murray~\cite{murray_heterogeneous_2011} suggests a value of 0.11 for the exclusive formation of $I_h$ observed in this
work. The variability of these experimental results stems mainly from the diversity of the kaolinite
samples (in terms of e.g. shape, purity and surfaces exposed, the latter still largely unknown) and the difficulty
to interpret the experimental data using heterogeneous CNT, for which tiny changes in quantities such as the free energy difference between water and
ice lead to substantial discrepancies~\cite{haji-akbari_direct_2015}. 
To date, experiments have to deal with populations of uneven particles and different nucleation sites.
Here we provide a value of $\mathcal{F}_S$ for a perfectly flat, defect-free (001) hydroxylated surface of kaolinite, in the hope to
aid the experimental investigation of well-defined, clean kaolinite substrates in the near future.
We also note that our simulations of crystal nucleation are the very edge of what molecular dynamics can presently achieve.
However, there is still room for improvement: for instance, heterogeneous ice formation can be affected by the presence of electric 
fields~\cite{ehre_water_2010,yan_heterogeneous_2011}, and similarly, water dissociation is common on many reactive surfaces~\cite{carrasco_molecular_2012}; 
these effects cannot be accounted for at present with the traditional force fields employed here.

In summary, we have calculated the heterogeneous ice nucleation rate for a fully atomistic water model on a prototypical
clay mineral of great importance to environmental science.  We have demonstrated that the hydroxylated (001) surface of
kaolinite boosts ice formation by twenty orders of magnitude with respect to homogeneous nucleation at the same
supercooling.  We have found that this particular kaolinite surface promotes the nucleation of the hexagonal ice
polytype, which forms thanks to the interaction of the prism face with the templating arrangements of hydroxyl groups at
the clay interface.  We have also found that ice nuclei tend to expand on the clay surface in two dimensions until they
reach the critical nucleus size. This is in contrast with the predictions of CNT, which however holds quantitatively for
ice formation on kaolinite even at this strong supercooling.  Finally, we provide a value of the heterogeneous shape
factor for the defect-free surface considered here, in the first attempt to bring simulations of heterogeneous ice
nucleation a step closer to experiments.  It remains to be investigated to what extent different surface morphologies
can in general affect nucleation rates or alter the ice polytypes which form. 

\section*{Acknowledgement}

This work was supported by the European Research Council under the European Union's Seventh Framework Programme
(FP/2007-2013)/ERC Grant Agreement number 616121 (HeteroIce project). A.M. is also supported by
a Royal Society Wolfson Research Merit Award. We are grateful for the computational resources provided
by the Swiss National Supercomputing Centre CSCS (Project s623 - Towards an Understanding of Ice Formation in Clouds).
T.L is supported by the National Science Fundation through the award CMMI-1537286.

%\bibliography{Text.bib}
%\input{Text.bbl}

\pagebreak
\widetext
\begin{center}
\textbf{\large SUPPORTING INFORMATION}
\end{center}
%%%%%%%%%% Merge with supplemental materials %%%%%%%%%%
%%%%%%%%%% Prefix a "S" to all equations, figures, tables and reset the counter %%%%%%%%%%
\setcounter{equation}{0}
\setcounter{figure}{0}
\setcounter{table}{0}
\setcounter{page}{1}
\makeatletter
\renewcommand{\theequation}{S\arabic{equation}}
\renewcommand{\thefigure}{S\arabic{figure}}
\renewcommand{\bibnumfmt}[1]{[S#1]}
\renewcommand{\citenumfont}[1]{S#1}
%%%%%%%%%% Prefix a "S" to all equations, figures, tables and reset the counter %%%%%%%%%%

We provide supporting information on the calculation of the heterogeneous ice nucleation rate on the kaolinite (001) hydroxylated surface.
The computational geometry is specified together with the details of the molecular dynamics simulations used in this work.
Moreover, we discuss the choice of the order parameter we have employed within the forward flux sampling calculations, and we provide additional
information about the implementation of the algorithm and the results obtained at each stage of the latter. A brief discussion about heterogeneous
classical nucleation theory is also presented together with the technical details of the topological criteria used to characterized the ice nuclei and 
a discussion about finite size effects.

\setcounter{figure}{0}
\begin{figure}[htbp]
\renewcommand\figurename{Fig.~S$\!\!$}
\centerline{\includegraphics[width=7cm]{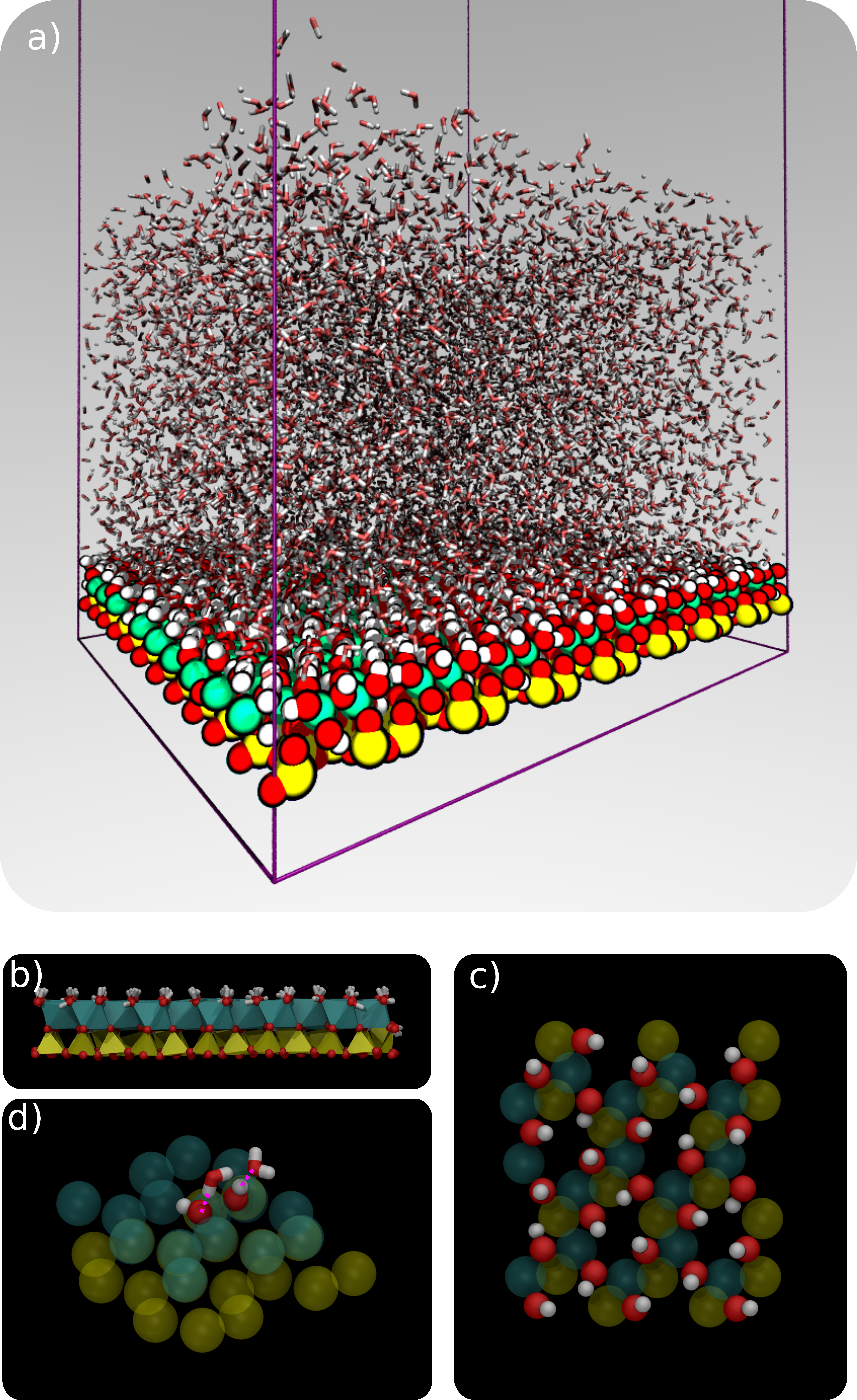}}
\caption{a) The simulation cell used in this work. A film of liquid water about 40 \AA$\ $ thick is in contact with a single slab of kaolinite,
 cleaved along the (001) plane. This slab geometry is thus characterized by two interfaces: the water-kaolinite interface and the water-vacuum interface.
The dimension of the simulation box along the normal to the slab is extended up to 150 \AA. Water molecules are depicted as sticks, while atoms
within the kaolinite slab as balls. Red, white, light blue and yellow atoms correspond to oxygen, hydrogen, aluminum and silicon atoms respectively.
b) (side view) The layered structure of the kaolinite slab: yellow tetrahedra and light blue octahedra represent the tetrahedral silica sheet
and the octahedral alumina sheet, terminated with hydroxyl groups, respectively. c) (top view) A small portion of the kaolinite slab depicting
the hexagonal arrangement of the hydroxyl groups exposed. d) Sketch of the amphoteric character of the hydroxylated (001) face of kaolinite: the
hydroxyl groups on top can either donate or accept an hydrogen bond from e.g. water molecules at the water-clay interface.}
\label{FIG_S1}
\end{figure}

%\begin{figure}[t!]
%\renewcommand\figurename{Fig.~S$\!\!$}
%\centerline{\includegraphics[width=7cm]{FIG_S2.pdf}}
%\caption{a) Probability density distribution for the order parameter $\lambda$ ($P(\lambda)$ left y-axis, boxes) and correspondent cumulative
%distribution function (CDF, right y-axes, empty circles). The blue and red vertical arrows mark the upper limit of the liquid basin $\lambda_{Liq}$
%and the position of $\lambda_0$ respectively. b) Flux rate ($\Phi_0$, left y-axis, filled circles) and number of direct crossing of the $\lambda_0$
%interface ($N_0$, right y-axis, empty circles) as a function of simulation time. c) Individual crossing probabilities $P(\lambda_i|\lambda_{i-1})$
%(normalized by their value at N=250) as a function of the number of crossing events.}
%\label{FIG_S2}
%\end{figure}

\section*{Computational Geometry}

The computational setup we have used is depicted in Fig.~S1a. A single layer of kaolinite, cleaved along the (001) plane
(perpendicular to the normal to the slab) was prepared by starting from the experimental cell parameters and
lattice positions~\cite{bish_rietveld_1993}. Specifically, a kaolinite bulk system made of two identical slabs was cleaved along the
(001) plane. The triclinc symmetry of the system (space group $C1$) was modified by setting the $\alpha$ and $\gamma$
angles (experimentally equal to 91.926 and 89.797 degrees respectively~\cite{bish_rietveld_1993}) to 90 degrees in order to make the cell
orthorombic. We explicitly verified that this modification does not introduce any structural change within the clay.
The final slab has in-plane dimensions of 61.84 and 71.54 \AA, corresponding to a 12 by 8 supercell.  We
positioned 6144 water molecules randomly atop this kaolinite slab at the density of the TIP4P/Ice model~\cite{abascal_potential_2005} at 300 K,
and expanded the dimension of the simulation cell along the normal to the slab to 150 \AA. This setup allows for a
physically meaningful equilibration of the water at the density of interest at a given temperature, but suffers from
two distinct drawbacks: i) the kaolinite slab possesses a net dipole moment which is not compensated throughout the simulation
cell and ii) the presence of the water-vacuum interface can alter the structure and the dynamics of the liquid film.
However, we have verified that compensating the dipole moment by means of a mirror slab does not affect our simulations,
as we have been able to replicate the results of Ref.~\citenum{zielke_simulations_2016} independently of the computational geometry. Furthermore, the water
film is thick enough to allow a bulk-like region to exist in terms of both structure and dynamics. The
effect of the water-vacuum interface is therefore negligible. In Fig.~S1b we highlight the layered nature of the slab, while in
Fig.~S1c we zoom in on a portion of the (001) hydroxylated surface and show the hexagonal arrangement of the hydroxyl groups.
This arrangement is important as the water can interact with the hydroxyls, so this arrangement is responsible for the templating effect of the clay 
which serves to promote ice nucleation. The amphoteric
nature of the hydroxyl groups at the surface is depicted in Fig.~S1d.

\section*{Molecular Dynamics Simulations}

The CLAY\_FF~\cite{cygan_molecular_2004} force field was used to model the kaolinite slab. We have not included the - optional - angular term (see
Ref.~\citenum{cygan_molecular_2004}), as we have verified that it does not affect the structure of the surface. In order to mimic the experimental
conditions, we have constrained the system at the experimental lateral dimensions (see above), and have also restrained the positions of the
silicon atoms at the bottom of the slab by means of an harmonic potential characterized by a spring constant of 1000
kJ/mol. All the other atoms within the kaolinite slab are unconstrained. We have verified that the thermal expansion of
the clay at 230 K ($\sim$ 0.4\% with respect to each lateral dimension) does not alter the structure nor the dynamics of the water-kaolinite interface. 
This setup is thus as close as we can get to the realistic (001) hydroxylated surface
within the CLAY\_FF model. The interaction between the water molecules have been modeled using the TIP4P/Ice model~\cite{abascal_potential_2005}, so
that our results are consistent with the homogeneous simulations of Ref.~\citenum{haji-akbari_direct_2015}. The interaction parameters between 
the clay and the water were obtained using the
standard Lorentz-Berthelot mixing rules~\cite{lorentz_ueber_1881,ahthefrench}.  Extreme care must be taken in order to correctly reproduce the
structure and the dynamics of the water-clay interface. The Forward Flux Sampling (FFS) simulations reported in this
work rely on a massive collection of unbiased Molecular Dynamics (MD) runs, all of which have been performed using the
GROMACS package, version 4.6.7. The code was compiled in single-precision, in order to alleviate the huge
computational workload needed to converge the FFS algorithm and because we have taken advantage of GPU acceleration, which is not
available in the double-precision version. The equations of motions were integrated using a leap-frog integrator
with a timestep of 2 fs. The van der Waals (non bonded) interactions were considered up to 10 \AA, where a
switching function was used to bring them to zero at 12 \AA.  Electrostatic interactions have been dealt with by means of an Ewald
summation up to 14 \AA. The NVT ensemble was sampled at 230 K using a stochastic velocity rescaling thermostat~\cite{bussi_canonical_2007} with
a very weak coupling constant of 4 ps in order to avoid temperature gradients throughout the system. The geometry
of the water molecules (TIP4P/Ice being a rigid model) was constrained using the SETTLE algorithm~\cite{miyamoto_settle:_1992} while the
P\_LINCS algorithm~\cite{hess_lincs:_1997} was used to constrain the O-H bonds within the clay.  We have verified that these settings reproduce
the dynamical properties of water reported in Ref.~\cite{haji-akbari_direct_2015}. The system was equilibrated at 300 K for 10 ns, before being
quenched to 230 K over 50 ns. This is the starting point for the calculation of the flux rate discussed in the next
section.

%\setcounter{figure}{0}
%\begin{figure}[b]
%\renewcommand\figurename{Fig.~S$\!\!$}
%\centerline{\includegraphics[width=7cm]{FIG_S1.pdf}}
%\caption{a) The simulation cell used in this work. A film of liquid water about 40 \AA$\ $ thick is in contact with a single slab of kaolinite,
% cleaved along the (001) plane. This slab geometry is thus characterized by two interfaces: the water-kaolinite interface and the water-vacuum interface.
%The dimension of the simulation box along the normal to the slab is extended up to 150 \AA. Water molecules are depicted as sticks, while atoms
%within the kaolinite slab as balls. Red, white, light blue and yellow atoms correspond to oxygen, hydrogen, aluminum and silicon atoms respectively.
%b) (side view) The layered structure of the kaolinite slab: yellow tetrahedra and light blue octahedra represent the tetrahedral silica sheet
%and the octahedral alumina sheet, terminated with hydroxyl groups, respectively. c) (top view) A small portion of the kaolinite slab depicting
%the hexagonal arrangement of the hydroxyl groups exposed. d) Sketch of the amphoteric character of the hydroxylated (001) face of kaolinite: the
%hydroxyl groups on top can either donate or accept an hydrogen bond from e.g. water molecules at the water-clay interface.}
%\label{FIG_S1}
%\end{figure}

\section*{Forward Flux Sampling Simulation}

\begin{figure}[t!]
\renewcommand\figurename{Fig.~S$\!\!$}
\centerline{\includegraphics[width=7cm]{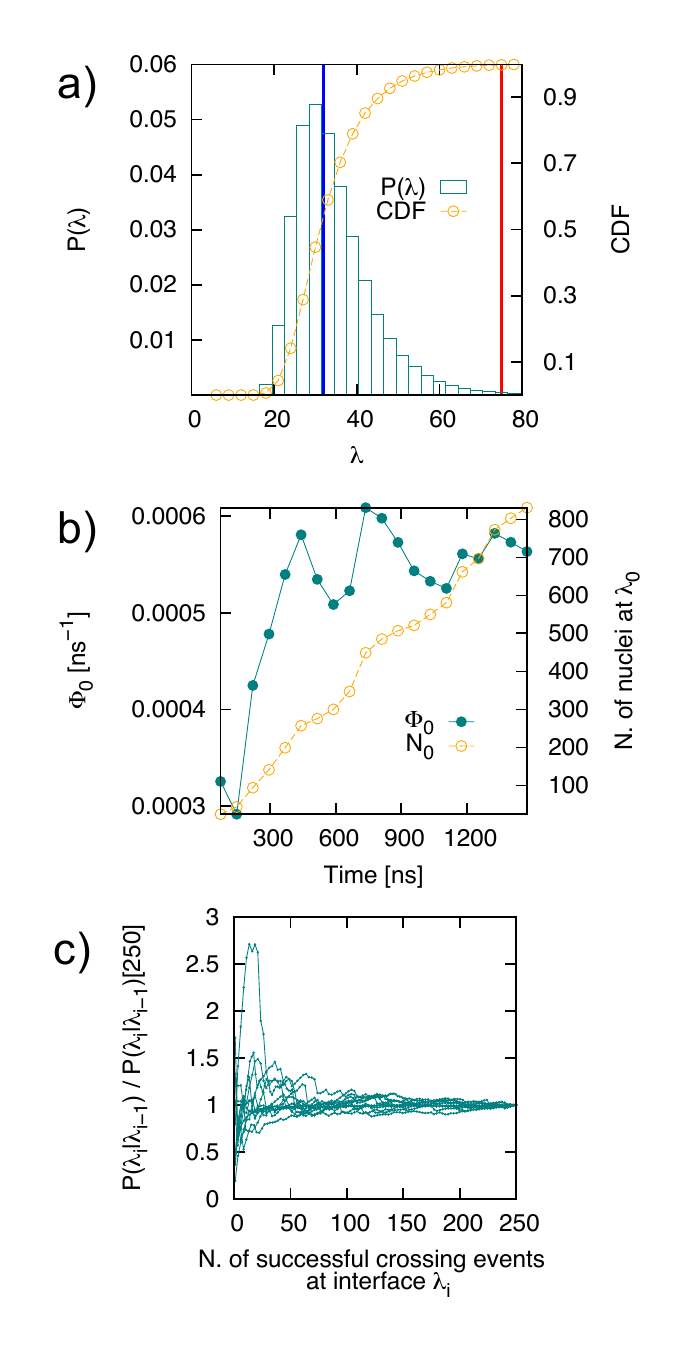}}
\caption{a) Probability density distribution for the order parameter $\lambda$ ($P(\lambda)$ left y-axis, boxes) and correspondent cumulative
distribution function (CDF, right y-axes, empty circles). The blue and red vertical arrows mark the upper limit of the liquid basin $\lambda_{Liq}$
and the position of $\lambda_0$ respectively. b) Flux rate ($\Phi_0$, left y-axis, filled circles) and number of direct crossing of the $\lambda_0$
interface ($N_0$, right y-axis, empty circles) as a function of simulation time. c) Individual crossing probabilities $P(\lambda_i|\lambda_{i-1})$
(normalized by their value at N=250) as a function of the number of crossing events.}
\label{FIG_S2}
\end{figure}

\subsection*{Order Parameter}

The first step in setting up the FFS simulation involved choosing a suitable order parameter $\lambda$. We start by
labeling as ice-like any water molecule whose oxygen atom displays a value of $lq^6$>0.45, where $lq^6$ is constructed as
follows: we first select only those oxygens which are hydrogen-bonded to four other oxygens.  For each of the $i-$th atoms of this
subset $S_{4HB}$, we calculate the local order parameter:

\begin{equation}
lq^6_i=\frac{\sum_{j=1}^{N_{S_{4HB}}} \sigma(\mathbf{r}_{ij}) \sum_{m=-6}^{6} q^{6*}_{i,m} \cdot q^6_{j,m} }  {\sum_{j=1}^{N_{S_{4HB}}} \sigma(\mathbf{r}_{ij}) }
\label{eq:L1}
\end{equation}

\noindent where $\sigma(\mathbf{r}_{ij})$ is a switching function tuned so that $\sigma(\mathbf{r}_{ij})$=1 when atom $j$ lies
within the first coordination shell of atom $i$ and which is zero otherwise. $q^6_{i,m}$ is the Steinhardt vector~\cite{steinhardt_bond-orientational_1983}

\begin{equation}
q^6_{i,m}=\frac { \sum_{j=1}^{N_{S_{4HB}}} \sigma(\mathbf{r}_{ij}) Y_{6m} (\mathbf{r}_{ij})  }    { \sum_{j=1}^{N_{S_{4HB}}} \sigma(\mathbf{r}_{ij})  },
\label{eq:L2}
\end{equation}

\noindent $Y_{6m} (\mathbf{r}_{ij})$ being one of the 6th order spherical harmonics. We have used 3.2 \AA$\ $ as the cutoff for
$\sigma(\mathbf{r}_{ij})$ to be consistent with Ref.~\citenum{haji-akbari_direct_2015}. Note that by selecting oxygen atoms within the
$S_{4HB}$ subset exclusively we ensure that the hydrogen bond network within the ice nuclei is reasonable. Having identified a set of ice-like water molecules, we pinpoint all the
connected clusters of oxygen atoms which: i) belong to the $S_{4HB}$ subset; ii) have a value of $lq^6$>0.45 and; iii)
are separated by a distance $\leq$ 3.2 \AA. We then select the largest of these clusters (i.e. the one
containing the largest number of oxygen atoms or equivalently water molecules). The final step is to find all the \textit{surface
molecules} that are connected to this cluster, as this procedure allows us to account for the diffuse interface between the solid and the liquid. 
Surface molecules are defined as the water molecules that lie within 3.2 \AA$\ $ from
the molecules in the cluster.
The final order parameter $\lambda$ used in this work is thus the number of water molecules
within the largest ice-like cluster plus the number of surface molecules. This approach allow us to include ice-like
atoms sitting directly on top of the kaolinite surface, which are never labeled as ice-like (and which would thus never be
included into the ice nuclei) because they are undercoordinated and because they display a different symmetry to
the molecules within bulk water (which in turn leads to different values of $lq^6$). 
Note that the order parameter used in Ref.~\citenum{haji-akbari_direct_2015} differs with respect to our formulation in that i) a slightly stricter criterion
has been used to label molecules as ice-like, namely $lq^6$>0.5 to be compared with our choice of $lq^6$>0.45; and ii) surface molecules are not included
in the largest ice-like nucleus. This means that in order to compare quantitatively our results with those of Ref.~\citenum{haji-akbari_direct_2015} in terms of e.g. 
the size of the critical nucleus, the very same order parameter has to be used.
The calculation of the order parameter is
performed on the fly during our MD simulations thanks to the flexibility of the PLUMED plugin~\cite{tribello_plumed_2014} (version 2.2).
This code deals chiefly with metadynamics simulations, but can be adapted to a FFS simulation. Note that PLUMED benefits
from a fully parallel implementation that flawlessly couples with the GPU-accelerated version of GROMACS, and thus provides
a very fast tool for performing FFS simulations.  Indeed, while several implementations of FFS are beginning to appear,
the main issue preventing wider adoption remains the implementation of the order parameter, which can be as complex as the one used in this work.
PLUMED allows a wide range of order parameters to be exploited without the need to re-code them elsewhere.

\subsection*{Converging the Flux Rate and the Individual Crossing Probabilities}

In order to calculate the flux rate ${\Phi_0}$ we have performed a 1.5 ms long unbiased MD simulation, and subsequently built
the probability density distribution for $P(\lambda)$ shown in Fig.~S2a. We have thus delimited the liquid basin in terms of the
order parameter as $0 < \lambda < \lambda_{Liq}=32$, while setting the initial interface for the FFS $\lambda_0$=75, corresponding
to a value of the cumulative distribution function of $P(\lambda)$ (also reported in Fig.~S2a) of 0.99. The flux rate is then computed
as the number of direct crossings of $\lambda_0$ (i.e. coming from $\lambda < \lambda_{Liq}$) divided by the total simulation time, and as
such should flatten as a function of time. Meanwhile, the number of direct crossings should increase linearly with time.
The value obtained for ${\Phi_0}$ and the number of crossings as a function of time are reported in Fig.~S2b. This figure demonstrates that, as
previously noted in Ref.~\citenum{bi_probing_2014}, long simulation times are needed in order to converge this quantity for inhomogeneous systems.
The calculated value of ${\Phi_0}$ is 0.00056359 ps$^{-1}$, which normalized by the average volume of the water film 
(189350.2980352 \AA$^3$) leads to the final value of 3.0$\cdot$10$^{-9 \pm 1}$ ps$^{-1}$ \AA$^{-3}$. Note that we have chosen to
normalize the flux rate by the average volume of the water film instead of by the surface area for the slab. While the latter choice could
in principle be thought as more meaningful in the context of heterogeneous nucleation, our objective is to compare our numbers with
the homogeneous case, which is why we choose the volume normalization rather than the surface area one. However, it should be noticed that
the two different normalizations only introduce a difference of an order of magnitude in the nucleation rate. The number of starting configurations, one
for each direct crossing of $\lambda_0$, is of the order of eight hundred, providing a comprehensive sampling including ice-like clusters
in the bulk of the water film as well as on top of the water surface (albeit the latter represent about 25\%).

Converging the individual crossing probabilities $P(\lambda_i|\lambda_{i-1})$ required in our case as many as 10,000
trial MD runs for the first few interfaces. The initial velocities for each MD run were randomly initialized
consistent with the corresponding Maxwell-Boltzmann distribution at 230 K. In line with the coarse graining
approach discussed in Ref.~\citenum{haji-akbari_direct_2015}, we have decided to compute the value of $\lambda$ on the
fly every 4 ps, a frequency far smaller than the relaxation time of the liquid at this temperature (about 0.5 ns) which
allows us to neglect meaningless fluctuation on very short timescales. The individual crossing probabilities, normalized
by their value after 250 crossing events, are reported in Fig.~S2c. Note that at the interfaces corresponding to
critical/post-critical ice nuclei a much smaller number (about 500) of trial MD runs have been shot, as for large ice
nuclei to get back to the liquid phase simulation times of the order of 10-40 ns are needed, dramatically increasing the
computational cost - albeit more and more nuclei proceed to grow as $\lambda$ increases leading to a faster convergence
of the crossing probabilities.  In fact, crossings for n>250 are not reported in Fig.~S2c as the crossing probabilities
are already converged well before n=250 within the last stages of the algorithm.
The confidence intervals for each $P(\lambda_i|\lambda_{i-1})$ have been computed according to the binomial distribution of 
the number of successful trial runs collected at $\lambda_i$ (see e.g. Ref.~\cite{allen_forward_2006}).

%\begin{figure}[t!]
%\renewcommand\figurename{Fig.~S$\!\!$}
%\centerline{\includegraphics[width=7cm]{FIG_S2.pdf}}
%\caption{a) Probability density distribution for the order parameter $\lambda$ ($P(\lambda)$ left y-axis, boxes) and correspondent cumulative
%distribution function (CDF, right y-axes, empty circles). The blue and red vertical arrows mark the upper limit of the liquid basin $\lambda_{Liq}$
%and the position of $\lambda_0$ respectively. b) Flux rate ($\Phi_0$, left y-axis, filled circles) and number of direct crossing of the $\lambda_0$
%interface ($N_0$, right y-axis, empty circles) as a function of simulation time. c) Individual crossing probabilities $P(\lambda_i|\lambda_{i-1})$
%(normalized by their value at N=250) as a function of the number of crossing events.}
%\label{FIG_S2}
%\end{figure}

\section*{Heterogeneous Classical Nucleation Theory}

Within the framework of classical nucleation theory, the homogeneous rate of nucleation $J_{Homo}$ can be written as~\cite{sear_nucleation:_2007,kalikmanov_nucleation_2013}:

\begin{equation}
J_{Homo}=A_{Homo}\cdot e^{- \frac {\Delta G^*_{Homo}} {k_B T} }
\end{equation}

\noindent where $A_{Homo}$ is a kinetic prefactor, $\Delta G^*_{Homo}$ is the height of the free energy barrier for nucleation and $k_B$ is the Boltzmann constant.
On the other hand, the heterogeneous rate of nucleation $J_{Hetero}$ can be written as~\cite{sear_nucleation:_2007,kalikmanov_nucleation_2013}:

\begin{equation}
J_{Hetero}=A_{Hetero}\cdot e^{- \frac {\mathcal{F}_S \cdot \Delta G^*_{Homo}} {k_B T} }
\end{equation}

\noindent where $A_{Hetero}$ is a kinetic prefactor which in principle can differ from $A_{Homo}$ and $\mathcal{F}_S$ is a shape factor,
or potency factor, which embeds the effectiveness of the substrate to promote nucleation. The value of $\mathcal{F}_S$
ranges from one (the surface does not contribute at all in lowering the free energy barrier for nucleation) to 0 (the
nucleation proceeds in a barrierless fashion). By taking the ratio $\frac{J_{Hetero}}{J_{Homo}}$ and assuming that
$A_{Hetero}=A_{Homo}$ (which is in many cases a perfectly reasonable assumption, see e.g.
Refs.~\citenum{li_homogeneous_2011,li_ice_2013,gianetti_computational_2016}), one can write the shape factor for
heterogeneous nucleation as:

\begin{equation}
\mathcal{F}_S=1-\left [\frac{k_B T}{\Delta G^*_{Homo}}\cdot \ln \left ( \frac{J_{Hetero}}{J_{Homo}}  \right ) \right ]
\end{equation}

The value of $\Delta G^*_{Homo}=\frac{1}{2}|\Delta\mu_{sl}| N^C_{Homo}$ is 80$\pm$ 5 $k_B T$, obtained from
Ref.~\citenum{haji-akbari_direct_2015} by using the definition of $\lambda$ we have employed here 
(thus using a slighlty different $lq_i^6$ cutoff and including surface molecules, see Eq.~\ref{eq:L1}) - which accounts for
an homogeneous critical nucleus size of 500 $\pm$ 30 water molecules and makes a direct comparison possible. 
Inserting this value into the expression above leads to a shape factor of $0.4\pm 0.1$.

\section*{Double-Diamond and Hexagonal Cages}

Double-Diamond (DDC) and Hexagonal cages (HC) are the building blocks of cubic and hexagonal ice respectively. We have identified water molecules involved in 
DDC and/or HC within the largest ice nucleus in the system (defined according to the order parameter $\lambda$, see Eqs. 1 and 2) following the
topological criteria detailed in Ref.~\citenum{haji-akbari_direct_2015}. The first step in order to locate DDC and HC is the construction of the ring network
of the oxygen atoms belonging to each water molecule. In this work, we have obtained all the six-atom rings needed to build DDC and HC using King's shortest
path criterion~\cite{king_ring_1967,franzblau_computation_1991} as implemented in the R.I.N.G.S. code~\cite{le_roux_ring_2010}. The same distance cutoff of 3.2 \AA$\ $ used 
for the construction of the order parameter $\lambda$ has been employed to determine the nearest neighbors of each oxygen atom. The same algorithm described in 
Ref.~\citenum{haji-akbari_direct_2015} has subsequently been used to determine DDC and HC.

\section*{Asphericity Parameter}

Many different choices are available to quantify the asphericity of clusters of molecules. We have considered the gyration radius as well as the $\alpha$ (
$\Delta$ in Ref.~\citenum{rawat_shape_2011}) and
$S$ asphericity parameters reported in Ref.~\citenum{rawat_shape_2011}. All of these quantities provided the same qualitative picture, so we have chosen to report
the asphericity trends for $\alpha$ only, the latter being defined as:

\begin{equation}
\alpha=\frac{3}{2 (tr\mathcal{T})^2} \sum_{i=1}^3 (\mu_i-\bar{\mu})^2
\end{equation}

where $\mu_i$ are the three eigenvalues of the inertia tensor $\mathcal{T}$ for a given cluster, and $\bar{\mu}=\frac{tr\mathcal{T}}{3}=\frac{\sum_{i=1}^3 (\mu_i)}{3}$ 

\section*{Spatial extent $\Delta z$}

The spatial extent $\Delta z$ for a given ice nucleus has been calculated as the difference between the minimum and maximum values of the z- components of the position vector of all
the oxygens belonging to the nucleus. As the direction normal to the kaolinite slab coincides to the z-axis of our simulation box, $\Delta z$ provides a qualitative indication of
the number of ice layers in the nuclei. Ice nuclei are defined to be on top of the kaolinite surface ($Surf$, see main text) if the minimum value of the z- components of the position vector of all
the oxygens belonging to the nucleus is < 15.0 \AA, which correspond to the position of the main peak in the density profile of the water film along the z-axis. If this is
not the case, the ice nuclei are considered to sit in the bulk of the water film ($Bulk$, see main text).

\section*{Avoiding Finite Size Effects}

Special care has to be taken when dealing with atomistic simulations of crystal nucleation from the liquid phase. Specifically, the presence
of periodic boundary conditions can introduce significant finite effects, most notably spurious interactions between the crystalline nuclei and their
periodic images. This artefact results in nonphysically large nucleation rates and/or crystal growth speeds. In this work we have considered
simulation boxes with lateral dimensions of the order of 60 \AA, which is sufficient to ensure that finite size effects do not affect our results.
We also measured the distance between the ice nuclei and their periodic images using the average set-set distance $d(A,B)$, which is defined as:

\begin{equation}
d(A,B)=\text{inf}\lim_{x\in A, y\in B} |\mathbf{x}-\mathbf{y}|
\end{equation}

\noindent where $x$ and $y$ are the position vectors of each oxygen atoms belonging to the largest ice nucleus (defined according to the order parameter 
$\lambda$) 
$A$ and its first periodic image $B$ respectively. At the FFS interface closest to the critical nucleus size                            
($\lambda$=225), $d(A,B)$=20$\pm$6\AA, and even at the last FFS interface we have considered ($\lambda$=325) the ice nuclei are still 
quite far away from their periodic images, $d(A,B)$ being 15$\pm$7\AA, which is of the order of 1/4 of the lateral dimension of the simulation box.

\providecommand{\latin}[1]{#1}
\providecommand*\mcitethebibliography{\thebibliography}
\csname @ifundefined\endcsname{endmcitethebibliography}
  {\let\endmcitethebibliography\endthebibliography}{}


\begin{thebibliography}{44}%
\makeatletter
\providecommand \@ifxundefined [1]{%
 \@ifx{#1\undefined}
}%
\providecommand \@ifnum [1]{%
 \ifnum #1\expandafter \@firstoftwo
 \else \expandafter \@secondoftwo
 \fi
}%
\providecommand \@ifx [1]{%
 \ifx #1\expandafter \@firstoftwo
 \else \expandafter \@secondoftwo
 \fi
}%
\providecommand \natexlab [1]{#1}%
\providecommand \enquote  [1]{``#1''}%
\providecommand \bibnamefont  [1]{#1}%
\providecommand \bibfnamefont [1]{#1}%
\providecommand \citenamefont [1]{#1}%
\providecommand \href@noop [0]{\@secondoftwo}%
\providecommand \href [0]{\begingroup \@sanitize@url \@href}%
\providecommand \@href[1]{\@@startlink{#1}\@@href}%
\providecommand \@@href[1]{\endgroup#1\@@endlink}%
\providecommand \@sanitize@url [0]{\catcode `\\12\catcode `\$12\catcode
  `\&12\catcode `\#12\catcode `\^12\catcode `\_12\catcode `\%12\relax}%
\providecommand \@@startlink[1]{}%
\providecommand \@@endlink[0]{}%
\providecommand \url  [0]{\begingroup\@sanitize@url \@url }%
\providecommand \@url [1]{\endgroup\@href {#1}{\urlprefix }}%
\providecommand \urlprefix  [0]{URL }%
\providecommand \Eprint [0]{\href }%
\providecommand \doibase [0]{http://dx.doi.org/}%
\providecommand \selectlanguage [0]{\@gobble}%
\providecommand \bibinfo  [0]{\@secondoftwo}%
\providecommand \bibfield  [0]{\@secondoftwo}%
\providecommand \translation [1]{[#1]}%
\providecommand \BibitemOpen [0]{}%
\providecommand \bibitemStop [0]{}%
\providecommand \bibitemNoStop [0]{.\EOS\space}%
\providecommand \EOS [0]{\spacefactor3000\relax}%
\providecommand \BibitemShut  [1]{\csname bibitem#1\endcsname}%
\let\auto@bib@innerbib\@empty
%</preamble>
\bibitem [{\citenamefont {Tam}\ \emph {et~al.}(2009)\citenamefont {Tam},
  \citenamefont {Rowley}, \citenamefont {Petrov}, \citenamefont {Zhang},
  \citenamefont {Afagh}, \citenamefont {Woo},\ and\ \citenamefont
  {Ben}}]{tam_solution_2009}%
  \BibitemOpen
  \bibfield  {author} {\bibinfo {author} {\bibfnamefont {R.~Y.}\ \bibnamefont
  {Tam}}, \bibinfo {author} {\bibfnamefont {C.~N.}\ \bibnamefont {Rowley}},
  \bibinfo {author} {\bibfnamefont {I.}~\bibnamefont {Petrov}}, \bibinfo
  {author} {\bibfnamefont {T.}~\bibnamefont {Zhang}}, \bibinfo {author}
  {\bibfnamefont {N.~A.}\ \bibnamefont {Afagh}}, \bibinfo {author}
  {\bibfnamefont {T.~K.}\ \bibnamefont {Woo}}, \ and\ \bibinfo {author}
  {\bibfnamefont {R.~N.}\ \bibnamefont {Ben}},\ }\href {\doibase
  10.1021/ja904169a} {\bibfield  {journal} {\bibinfo  {journal} {J. Am. Chem.
  Soc.}\ }\textbf {\bibinfo {volume} {131}},\ \bibinfo {pages} {15745}
  (\bibinfo {year} {2009})}\BibitemShut {NoStop}%
\bibitem [{\citenamefont {Bolton}\ and\ \citenamefont
  {Pettersson}(2001)}]{bolton_ice-catalyzed_2001}%
  \BibitemOpen
  \bibfield  {author} {\bibinfo {author} {\bibfnamefont {K.}~\bibnamefont
  {Bolton}}\ and\ \bibinfo {author} {\bibfnamefont {J.~B.~C.}\ \bibnamefont
  {Pettersson}},\ }\href {\doibase 10.1021/ja010096c} {\bibfield  {journal}
  {\bibinfo  {journal} {J. Am. Chem. Soc.}\ }\textbf {\bibinfo {volume}
  {123}},\ \bibinfo {pages} {7360} (\bibinfo {year} {2001})}\BibitemShut
  {NoStop}%
\bibitem [{\citenamefont {Tang}, \citenamefont {Cziczo},\ and\ \citenamefont
  {Grassian}(2016)}]{tang_interactions_2016}%
  \BibitemOpen
  \bibfield  {author} {\bibinfo {author} {\bibfnamefont {M.}~\bibnamefont
  {Tang}}, \bibinfo {author} {\bibfnamefont {D.~J.}\ \bibnamefont {Cziczo}}, \
  and\ \bibinfo {author} {\bibfnamefont {V.~H.}\ \bibnamefont {Grassian}},\
  }\href {\doibase 10.1021/acs.chemrev.5b00529} {\bibfield  {journal} {\bibinfo
   {journal} {Chem. Rev.}\ }\textbf {\bibinfo {volume} {116}},\ \bibinfo
  {pages} {4205} (\bibinfo {year} {2016})}\BibitemShut {NoStop}%
\bibitem [{\citenamefont {Bartels-Rausch}\ \emph {et~al.}(2012)\citenamefont
  {Bartels-Rausch}, \citenamefont {Bergeron}, \citenamefont {Cartwright},
  \citenamefont {Escribano}, \citenamefont {Finney}, \citenamefont {Grothe},
  \citenamefont {Gutiérrez}, \citenamefont {Haapala}, \citenamefont {Kuhs},
  \citenamefont {Pettersson}, \citenamefont {Price}, \citenamefont
  {Sainz-Díaz}, \citenamefont {Stokes}, \citenamefont {Strazzulla},
  \citenamefont {Thomson}, \citenamefont {Trinks},\ and\ \citenamefont
  {Uras-Aytemiz}}]{bartels-rausch_ice_2012}%
  \BibitemOpen
  \bibfield  {author} {\bibinfo {author} {\bibfnamefont {T.}~\bibnamefont
  {Bartels-Rausch}}, \bibinfo {author} {\bibfnamefont {V.}~\bibnamefont
  {Bergeron}}, \bibinfo {author} {\bibfnamefont {J.~H.~E.}\ \bibnamefont
  {Cartwright}}, \bibinfo {author} {\bibfnamefont {R.}~\bibnamefont
  {Escribano}}, \bibinfo {author} {\bibfnamefont {J.~L.}\ \bibnamefont
  {Finney}}, \bibinfo {author} {\bibfnamefont {H.}~\bibnamefont {Grothe}},
  \bibinfo {author} {\bibfnamefont {P.~J.}\ \bibnamefont {Gutiérrez}},
  \bibinfo {author} {\bibfnamefont {J.}~\bibnamefont {Haapala}}, \bibinfo
  {author} {\bibfnamefont {W.~F.}\ \bibnamefont {Kuhs}}, \bibinfo {author}
  {\bibfnamefont {J.~B.~C.}\ \bibnamefont {Pettersson}}, \bibinfo {author}
  {\bibfnamefont {S.~D.}\ \bibnamefont {Price}}, \bibinfo {author}
  {\bibfnamefont {C.~I.}\ \bibnamefont {Sainz-Díaz}}, \bibinfo {author}
  {\bibfnamefont {D.~J.}\ \bibnamefont {Stokes}}, \bibinfo {author}
  {\bibfnamefont {G.}~\bibnamefont {Strazzulla}}, \bibinfo {author}
  {\bibfnamefont {E.~S.}\ \bibnamefont {Thomson}}, \bibinfo {author}
  {\bibfnamefont {H.}~\bibnamefont {Trinks}}, \ and\ \bibinfo {author}
  {\bibfnamefont {N.}~\bibnamefont {Uras-Aytemiz}},\ }\href {\doibase
  10.1103/RevModPhys.84.885} {\bibfield  {journal} {\bibinfo  {journal} {Rev.
  Mod. Phys.}\ }\textbf {\bibinfo {volume} {84}},\ \bibinfo {pages} {885}
  (\bibinfo {year} {2012})}\BibitemShut {NoStop}%
\bibitem [{\citenamefont {Pirzadeh}\ and\ \citenamefont
  {Kusalik}(2013)}]{pirzadeh_molecular_2013}%
  \BibitemOpen
  \bibfield  {author} {\bibinfo {author} {\bibfnamefont {P.}~\bibnamefont
  {Pirzadeh}}\ and\ \bibinfo {author} {\bibfnamefont {P.~G.}\ \bibnamefont
  {Kusalik}},\ }\href {\doibase 10.1021/ja400521e} {\bibfield  {journal}
  {\bibinfo  {journal} {J. Am. Chem. Soc.}\ }\textbf {\bibinfo {volume}
  {135}},\ \bibinfo {pages} {7278} (\bibinfo {year} {2013})}\BibitemShut
  {NoStop}%
\bibitem [{\citenamefont {Murray}\ \emph {et~al.}(2012)\citenamefont {Murray},
  \citenamefont {O'Sullivan}, \citenamefont {Atkinson},\ and\ \citenamefont
  {Webb}}]{murray_ice_2012}%
  \BibitemOpen
  \bibfield  {author} {\bibinfo {author} {\bibfnamefont {B.~J.}\ \bibnamefont
  {Murray}}, \bibinfo {author} {\bibfnamefont {D.}~\bibnamefont {O'Sullivan}},
  \bibinfo {author} {\bibfnamefont {J.~D.}\ \bibnamefont {Atkinson}}, \ and\
  \bibinfo {author} {\bibfnamefont {M.~E.}\ \bibnamefont {Webb}},\ }\href
  {\doibase 10.1039/C2CS35200A} {\bibfield  {journal} {\bibinfo  {journal}
  {Chem. Soc. Rev.}\ }\textbf {\bibinfo {volume} {41}},\ \bibinfo {pages}
  {6519} (\bibinfo {year} {2012})}\BibitemShut {NoStop}%
\bibitem [{\citenamefont {Lupi}, \citenamefont {Hudait},\ and\ \citenamefont
  {Molinero}(2014)}]{lupi_heterogeneous_2014}%
  \BibitemOpen
  \bibfield  {author} {\bibinfo {author} {\bibfnamefont {L.}~\bibnamefont
  {Lupi}}, \bibinfo {author} {\bibfnamefont {A.}~\bibnamefont {Hudait}}, \ and\
  \bibinfo {author} {\bibfnamefont {V.}~\bibnamefont {Molinero}},\ }\href
  {\doibase 10.1021/ja411507a} {\bibfield  {journal} {\bibinfo  {journal} {J.
  Am. Chem. Soc.}\ }\textbf {\bibinfo {volume} {136}},\ \bibinfo {pages} {3156}
  (\bibinfo {year} {2014})}\BibitemShut {NoStop}%
\bibitem [{\citenamefont {O'Sullivan}\ \emph {et~al.}(2015)\citenamefont
  {O'Sullivan}, \citenamefont {Murray}, \citenamefont {Ross}, \citenamefont
  {Whale}, \citenamefont {Price}, \citenamefont {Atkinson}, \citenamefont
  {Umo},\ and\ \citenamefont {Webb}}]{osullivan_relevance_2015}%
  \BibitemOpen
  \bibfield  {author} {\bibinfo {author} {\bibfnamefont {D.}~\bibnamefont
  {O'Sullivan}}, \bibinfo {author} {\bibfnamefont {B.~J.}\ \bibnamefont
  {Murray}}, \bibinfo {author} {\bibfnamefont {J.~F.}\ \bibnamefont {Ross}},
  \bibinfo {author} {\bibfnamefont {T.~F.}\ \bibnamefont {Whale}}, \bibinfo
  {author} {\bibfnamefont {H.~C.}\ \bibnamefont {Price}}, \bibinfo {author}
  {\bibfnamefont {J.~D.}\ \bibnamefont {Atkinson}}, \bibinfo {author}
  {\bibfnamefont {N.~S.}\ \bibnamefont {Umo}}, \ and\ \bibinfo {author}
  {\bibfnamefont {M.~E.}\ \bibnamefont {Webb}},\ }\href {\doibase
  10.1038/srep08082} {\bibfield  {journal} {\bibinfo  {journal} {Sci. Rep.}\
  }\textbf {\bibinfo {volume} {5}},\ \bibinfo {pages} {1} (\bibinfo {year}
  {2015})}\BibitemShut {NoStop}%
\bibitem [{\citenamefont {Eastwood}\ \emph {et~al.}(2008)\citenamefont
  {Eastwood}, \citenamefont {Cremel}, \citenamefont {Gehrke}, \citenamefont
  {Girard},\ and\ \citenamefont {Bertram}}]{eastwood_ice_2008}%
  \BibitemOpen
  \bibfield  {author} {\bibinfo {author} {\bibfnamefont {M.~L.}\ \bibnamefont
  {Eastwood}}, \bibinfo {author} {\bibfnamefont {S.}~\bibnamefont {Cremel}},
  \bibinfo {author} {\bibfnamefont {C.}~\bibnamefont {Gehrke}}, \bibinfo
  {author} {\bibfnamefont {E.}~\bibnamefont {Girard}}, \ and\ \bibinfo {author}
  {\bibfnamefont {A.~K.}\ \bibnamefont {Bertram}},\ }\href {\doibase
  10.1029/2008JD010639} {\bibfield  {journal} {\bibinfo  {journal} {J. Geophys.
  Res.}\ }\textbf {\bibinfo {volume} {113}},\ \bibinfo {pages} {D22203}
  (\bibinfo {year} {2008})}\BibitemShut {NoStop}%
\bibitem [{\citenamefont {Fitzner}\ \emph {et~al.}(2015)\citenamefont
  {Fitzner}, \citenamefont {Sosso}, \citenamefont {Cox},\ and\ \citenamefont
  {Michaelides}}]{fitzner_many_2015}%
  \BibitemOpen
  \bibfield  {author} {\bibinfo {author} {\bibfnamefont {M.}~\bibnamefont
  {Fitzner}}, \bibinfo {author} {\bibfnamefont {G.~C.}\ \bibnamefont {Sosso}},
  \bibinfo {author} {\bibfnamefont {S.~J.}\ \bibnamefont {Cox}}, \ and\
  \bibinfo {author} {\bibfnamefont {A.}~\bibnamefont {Michaelides}},\ }\href
  {\doibase 10.1021/jacs.5b08748} {\bibfield  {journal} {\bibinfo  {journal}
  {J. Am. Chem. Soc.}\ }\textbf {\bibinfo {volume} {137}},\ \bibinfo {pages}
  {13658} (\bibinfo {year} {2015})}\BibitemShut {NoStop}%
\bibitem [{\citenamefont {Cabriolu}\ and\ \citenamefont
  {Li}(2015)}]{cabriolu_ice_2015}%
  \BibitemOpen
  \bibfield  {author} {\bibinfo {author} {\bibfnamefont {R.}~\bibnamefont
  {Cabriolu}}\ and\ \bibinfo {author} {\bibfnamefont {T.}~\bibnamefont {Li}},\
  }\href {\doibase 10.1103/PhysRevE.91.052402} {\bibfield  {journal} {\bibinfo
  {journal} {Phys. Rev. E}\ }\textbf {\bibinfo {volume} {91}},\ \bibinfo
  {pages} {052402} (\bibinfo {year} {2015})}\BibitemShut {NoStop}%
\bibitem [{\citenamefont {Zimmermann}\ \emph {et~al.}(2008)\citenamefont
  {Zimmermann}, \citenamefont {Weinbruch}, \citenamefont {Schi\"{u}tz},
  \citenamefont {Hofmann}, \citenamefont {Ebert}, \citenamefont {Kandler},\
  and\ \citenamefont {Worringen}}]{zimmermann_ice_2008}%
  \BibitemOpen
  \bibfield  {author} {\bibinfo {author} {\bibfnamefont {F.}~\bibnamefont
  {Zimmermann}}, \bibinfo {author} {\bibfnamefont {S.}~\bibnamefont
  {Weinbruch}}, \bibinfo {author} {\bibfnamefont {L.}~\bibnamefont
  {Schi\"{u}tz}}, \bibinfo {author} {\bibfnamefont {H.}~\bibnamefont
  {Hofmann}}, \bibinfo {author} {\bibfnamefont {M.}~\bibnamefont {Ebert}},
  \bibinfo {author} {\bibfnamefont {K.}~\bibnamefont {Kandler}}, \ and\
  \bibinfo {author} {\bibfnamefont {A.}~\bibnamefont {Worringen}},\ }\href
  {\doibase 10.1029/2008JD010655} {\bibfield  {journal} {\bibinfo  {journal}
  {J. Geophys. Res.}\ }\textbf {\bibinfo {volume} {113}},\ \bibinfo {pages}
  {D23204} (\bibinfo {year} {2008})}\BibitemShut {NoStop}%
\bibitem [{\citenamefont {Murray}\ \emph {et~al.}(2011)\citenamefont {Murray},
  \citenamefont {Broadley}, \citenamefont {Wilson}, \citenamefont {Atkinson},\
  and\ \citenamefont {Wills}}]{murray_heterogeneous_2011}%
  \BibitemOpen
  \bibfield  {author} {\bibinfo {author} {\bibfnamefont {B.~J.}\ \bibnamefont
  {Murray}}, \bibinfo {author} {\bibfnamefont {S.~L.}\ \bibnamefont
  {Broadley}}, \bibinfo {author} {\bibfnamefont {T.~W.}\ \bibnamefont
  {Wilson}}, \bibinfo {author} {\bibfnamefont {J.~D.}\ \bibnamefont
  {Atkinson}}, \ and\ \bibinfo {author} {\bibfnamefont {R.~H.}\ \bibnamefont
  {Wills}},\ }\href {\doibase 10.5194/acp-11-4191-2011} {\bibfield  {journal}
  {\bibinfo  {journal} {Atmos. Chem. Phys.}\ }\textbf {\bibinfo {volume}
  {11}},\ \bibinfo {pages} {4191} (\bibinfo {year} {2011})}\BibitemShut
  {NoStop}%
\bibitem [{\citenamefont {Tobo}\ \emph {et~al.}(2012)\citenamefont {Tobo},
  \citenamefont {DeMott}, \citenamefont {Raddatz}, \citenamefont {Niedermeier},
  \citenamefont {Hartmann}, \citenamefont {Kreidenweis}, \citenamefont
  {Stratmann},\ and\ \citenamefont {Wex}}]{tobo_impacts_2012}%
  \BibitemOpen
  \bibfield  {author} {\bibinfo {author} {\bibfnamefont {Y.}~\bibnamefont
  {Tobo}}, \bibinfo {author} {\bibfnamefont {P.~J.}\ \bibnamefont {DeMott}},
  \bibinfo {author} {\bibfnamefont {M.}~\bibnamefont {Raddatz}}, \bibinfo
  {author} {\bibfnamefont {D.}~\bibnamefont {Niedermeier}}, \bibinfo {author}
  {\bibfnamefont {S.}~\bibnamefont {Hartmann}}, \bibinfo {author}
  {\bibfnamefont {S.~M.}\ \bibnamefont {Kreidenweis}}, \bibinfo {author}
  {\bibfnamefont {F.}~\bibnamefont {Stratmann}}, \ and\ \bibinfo {author}
  {\bibfnamefont {H.}~\bibnamefont {Wex}},\ }\href {\doibase
  10.1029/2012GL053007} {\bibfield  {journal} {\bibinfo  {journal} {Geophys.
  Res. Lett.}\ }\textbf {\bibinfo {volume} {39}},\ \bibinfo {pages} {L19803}
  (\bibinfo {year} {2012})}\BibitemShut {NoStop}%
\bibitem [{\citenamefont {Welti}\ \emph {et~al.}(2013)\citenamefont {Welti},
  \citenamefont {Kanji}, \citenamefont {L\"{u}\"{o}d}, \citenamefont
  {Stetzer},\ and\ \citenamefont {Lohmann}}]{welti_exploring_2013}%
  \BibitemOpen
  \bibfield  {author} {\bibinfo {author} {\bibfnamefont {A.}~\bibnamefont
  {Welti}}, \bibinfo {author} {\bibfnamefont {Z.~A.}\ \bibnamefont {Kanji}},
  \bibinfo {author} {\bibfnamefont {F.}~\bibnamefont {L\"{u}\"{o}d}}, \bibinfo
  {author} {\bibfnamefont {O.}~\bibnamefont {Stetzer}}, \ and\ \bibinfo
  {author} {\bibfnamefont {U.}~\bibnamefont {Lohmann}},\ }\href {\doibase
  10.1175/JAS-D-12-0252.1} {\bibfield  {journal} {\bibinfo  {journal} {J.
  Atmos. Sci.}\ }\textbf {\bibinfo {volume} {71}},\ \bibinfo {pages} {16}
  (\bibinfo {year} {2013})}\BibitemShut {NoStop}%
\bibitem [{\citenamefont {Wex}\ \emph {et~al.}(2014)\citenamefont {Wex},
  \citenamefont {DeMott}, \citenamefont {Tobo}, \citenamefont {Hartmann},
  \citenamefont {R\"{o}sch}, \citenamefont {Clauss}, \citenamefont {Tomsche},
  \citenamefont {Niedermeier},\ and\ \citenamefont
  {Stratmann}}]{wex_kaolinite_2014}%
  \BibitemOpen
  \bibfield  {author} {\bibinfo {author} {\bibfnamefont {H.}~\bibnamefont
  {Wex}}, \bibinfo {author} {\bibfnamefont {P.~J.}\ \bibnamefont {DeMott}},
  \bibinfo {author} {\bibfnamefont {Y.}~\bibnamefont {Tobo}}, \bibinfo {author}
  {\bibfnamefont {S.}~\bibnamefont {Hartmann}}, \bibinfo {author}
  {\bibfnamefont {M.}~\bibnamefont {R\"{o}sch}}, \bibinfo {author}
  {\bibfnamefont {T.}~\bibnamefont {Clauss}}, \bibinfo {author} {\bibfnamefont
  {L.}~\bibnamefont {Tomsche}}, \bibinfo {author} {\bibfnamefont
  {D.}~\bibnamefont {Niedermeier}}, \ and\ \bibinfo {author} {\bibfnamefont
  {F.}~\bibnamefont {Stratmann}},\ }\href {\doibase 10.5194/acp-14-5529-2014}
  {\bibfield  {journal} {\bibinfo  {journal} {Atmos. Chem. Phys.}\ }\textbf
  {\bibinfo {volume} {14}},\ \bibinfo {pages} {5529} (\bibinfo {year}
  {2014})}\BibitemShut {NoStop}%
\bibitem [{\citenamefont {Hu}\ and\ \citenamefont
  {Michaelides}(2010)}]{hu_kaolinite_2010}%
  \BibitemOpen
  \bibfield  {author} {\bibinfo {author} {\bibfnamefont {X.~L.}\ \bibnamefont
  {Hu}}\ and\ \bibinfo {author} {\bibfnamefont {A.}~\bibnamefont
  {Michaelides}},\ }\href {\doibase 10.1016/j.susc.2009.10.026} {\bibfield
  {journal} {\bibinfo  {journal} {Surf. Sci.}\ }\textbf {\bibinfo {volume}
  {604}},\ \bibinfo {pages} {111} (\bibinfo {year} {2010})}\BibitemShut
  {NoStop}%
\bibitem [{\citenamefont {Tunega}, \citenamefont {Gerzabek},\ and\
  \citenamefont {Lischka}(2004)}]{tunega_ab_2004}%
  \BibitemOpen
  \bibfield  {author} {\bibinfo {author} {\bibfnamefont {D.}~\bibnamefont
  {Tunega}}, \bibinfo {author} {\bibfnamefont {M.~H.}\ \bibnamefont
  {Gerzabek}}, \ and\ \bibinfo {author} {\bibfnamefont {H.}~\bibnamefont
  {Lischka}},\ }\href {\doibase 10.1021/jp037121g} {\bibfield  {journal}
  {\bibinfo  {journal} {J. Phys. Chem. B}\ }\textbf {\bibinfo {volume} {108}},\
  \bibinfo {pages} {5930} (\bibinfo {year} {2004})}\BibitemShut {NoStop}%
\bibitem [{\citenamefont {Cox}\ \emph {et~al.}(2014)\citenamefont {Cox},
  \citenamefont {Raza}, \citenamefont {Kathmann}, \citenamefont {Slater},\ and\
  \citenamefont {Michaelides}}]{cox_microscopic_2014}%
  \BibitemOpen
  \bibfield  {author} {\bibinfo {author} {\bibfnamefont {S.~J.}\ \bibnamefont
  {Cox}}, \bibinfo {author} {\bibfnamefont {Z.}~\bibnamefont {Raza}}, \bibinfo
  {author} {\bibfnamefont {S.~M.}\ \bibnamefont {Kathmann}}, \bibinfo {author}
  {\bibfnamefont {B.}~\bibnamefont {Slater}}, \ and\ \bibinfo {author}
  {\bibfnamefont {A.}~\bibnamefont {Michaelides}},\ }\href {\doibase
  10.1039/C3FD00059A} {\bibfield  {journal} {\bibinfo  {journal} {Farad.
  Discuss.}\ }\textbf {\bibinfo {volume} {167}},\ \bibinfo {pages} {389}
  (\bibinfo {year} {2014})}\BibitemShut {NoStop}%
\bibitem [{\citenamefont {Zielke}, \citenamefont {Bertram},\ and\ \citenamefont
  {Patey}(2016)}]{zielke_simulations_2016}%
  \BibitemOpen
  \bibfield  {author} {\bibinfo {author} {\bibfnamefont {S.~A.}\ \bibnamefont
  {Zielke}}, \bibinfo {author} {\bibfnamefont {A.~K.}\ \bibnamefont {Bertram}},
  \ and\ \bibinfo {author} {\bibfnamefont {G.~N.}\ \bibnamefont {Patey}},\
  }\href {\doibase 10.1021/acs.jpcb.5b09052} {\bibfield  {journal} {\bibinfo
  {journal} {J. Phys. Chem. B}\ }\textbf {\bibinfo {volume} {120}},\ \bibinfo
  {pages} {1726} (\bibinfo {year} {2016})}\BibitemShut {NoStop}%
\bibitem [{\citenamefont {Sanz}\ \emph {et~al.}(2013)\citenamefont {Sanz},
  \citenamefont {Vega}, \citenamefont {Espinosa}, \citenamefont
  {Caballero-Bernal}, \citenamefont {Abascal},\ and\ \citenamefont
  {Valeriani}}]{sanz_homogeneous_2013}%
  \BibitemOpen
  \bibfield  {author} {\bibinfo {author} {\bibfnamefont {E.}~\bibnamefont
  {Sanz}}, \bibinfo {author} {\bibfnamefont {C.}~\bibnamefont {Vega}}, \bibinfo
  {author} {\bibfnamefont {J.~R.}\ \bibnamefont {Espinosa}}, \bibinfo {author}
  {\bibfnamefont {R.}~\bibnamefont {Caballero-Bernal}}, \bibinfo {author}
  {\bibfnamefont {J.~L.~F.}\ \bibnamefont {Abascal}}, \ and\ \bibinfo {author}
  {\bibfnamefont {C.}~\bibnamefont {Valeriani}},\ }\href {\doibase
  10.1021/ja4028814} {\bibfield  {journal} {\bibinfo  {journal} {J. Am. Chem.
  Soc.}\ }\textbf {\bibinfo {volume} {135}},\ \bibinfo {pages} {15008}
  (\bibinfo {year} {2013})}\BibitemShut {NoStop}%
\bibitem [{\citenamefont {Haji-Akbari}\ and\ \citenamefont
  {Debenedetti}(2015)}]{haji-akbari_direct_2015}%
  \BibitemOpen
  \bibfield  {author} {\bibinfo {author} {\bibfnamefont {A.}~\bibnamefont
  {Haji-Akbari}}\ and\ \bibinfo {author} {\bibfnamefont {P.~G.}\ \bibnamefont
  {Debenedetti}},\ }\href {\doibase 10.1073/pnas.1509267112} {\bibfield
  {journal} {\bibinfo  {journal} {Proc. Natl. Acad. Sci.}\ }\textbf {\bibinfo
  {volume} {112}},\ \bibinfo {pages} {10582} (\bibinfo {year}
  {2015})}\BibitemShut {NoStop}%
\bibitem [{\citenamefont {Pruppacher}\ and\ \citenamefont
  {Klett}(1997)}]{pruppacher_microphysics_1997}%
  \BibitemOpen
  \bibfield  {author} {\bibinfo {author} {\bibfnamefont {H.~R.}\ \bibnamefont
  {Pruppacher}}\ and\ \bibinfo {author} {\bibfnamefont {J.~D.}\ \bibnamefont
  {Klett}},\ }\href@noop {} {\emph {\bibinfo {title} {Microphysics Of Clouds
  And Precipitation}}}\ (\bibinfo  {publisher} {Springer Science \& Business
  Media},\ \bibinfo {year} {1997})\BibitemShut {NoStop}%
\bibitem [{\citenamefont {Abascal}\ \emph {et~al.}(2005)\citenamefont
  {Abascal}, \citenamefont {Sanz}, \citenamefont {Fernandez},\ and\
  \citenamefont {Vega}}]{abascal_potential_2005}%
  \BibitemOpen
  \bibfield  {author} {\bibinfo {author} {\bibfnamefont {J.~L.~F.}\
  \bibnamefont {Abascal}}, \bibinfo {author} {\bibfnamefont {E.}~\bibnamefont
  {Sanz}}, \bibinfo {author} {\bibfnamefont {R.~G.}\ \bibnamefont {Fernandez}},
  \ and\ \bibinfo {author} {\bibfnamefont {C.}~\bibnamefont {Vega}},\ }\href
  {\doibase 10.1063/1.1931662} {\bibfield  {journal} {\bibinfo  {journal} {J.
  Chem. Phys.}\ }\textbf {\bibinfo {volume} {122}},\ \bibinfo {pages} {234511}
  (\bibinfo {year} {2005})}\BibitemShut {NoStop}%
\bibitem [{\citenamefont {Hu}\ and\ \citenamefont
  {Michaelides}(2008)}]{hu_water_2008}%
  \BibitemOpen
  \bibfield  {author} {\bibinfo {author} {\bibfnamefont {X.~L.}\ \bibnamefont
  {Hu}}\ and\ \bibinfo {author} {\bibfnamefont {A.}~\bibnamefont
  {Michaelides}},\ }\href {\doibase 10.1016/j.susc.2007.12.032} {\bibfield
  {journal} {\bibinfo  {journal} {Surf. Sci.}\ }\textbf {\bibinfo {volume}
  {602}},\ \bibinfo {pages} {960} (\bibinfo {year} {2008})}\BibitemShut
  {NoStop}%
\bibitem [{\citenamefont {Allen}, \citenamefont {Valeriani},\ and\
  \citenamefont {Rein~ten Wolde}(2009)}]{allen_forward_2009}%
  \BibitemOpen
  \bibfield  {author} {\bibinfo {author} {\bibfnamefont {R.~J.}\ \bibnamefont
  {Allen}}, \bibinfo {author} {\bibfnamefont {C.}~\bibnamefont {Valeriani}}, \
  and\ \bibinfo {author} {\bibfnamefont {P.}~\bibnamefont {Rein~ten Wolde}},\
  }\href {\doibase 10.1088/0953-8984/21/46/463102} {\bibfield  {journal}
  {\bibinfo  {journal} {J. Phys. Cond. Matt.}\ }\textbf {\bibinfo {volume}
  {21}},\ \bibinfo {pages} {463102} (\bibinfo {year} {2009})}\BibitemShut
  {NoStop}%
\bibitem [{\citenamefont {Li}\ \emph {et~al.}(2011)\citenamefont {Li},
  \citenamefont {Donadio}, \citenamefont {Russo},\ and\ \citenamefont
  {Galli}}]{li_homogeneous_2011}%
  \BibitemOpen
  \bibfield  {author} {\bibinfo {author} {\bibfnamefont {T.}~\bibnamefont
  {Li}}, \bibinfo {author} {\bibfnamefont {D.}~\bibnamefont {Donadio}},
  \bibinfo {author} {\bibfnamefont {G.}~\bibnamefont {Russo}}, \ and\ \bibinfo
  {author} {\bibfnamefont {G.}~\bibnamefont {Galli}},\ }\href {\doibase
  10.1039/C1CP22167A} {\bibfield  {journal} {\bibinfo  {journal} {Phys. Chem.
  Chem. Phys.}\ }\textbf {\bibinfo {volume} {13}},\ \bibinfo {pages} {19807}
  (\bibinfo {year} {2011})}\BibitemShut {NoStop}%
\bibitem [{\citenamefont {Li}, \citenamefont {Donadio},\ and\ \citenamefont
  {Galli}(2013)}]{li_ice_2013}%
  \BibitemOpen
  \bibfield  {author} {\bibinfo {author} {\bibfnamefont {T.}~\bibnamefont
  {Li}}, \bibinfo {author} {\bibfnamefont {D.}~\bibnamefont {Donadio}}, \ and\
  \bibinfo {author} {\bibfnamefont {G.}~\bibnamefont {Galli}},\ }\href
  {\doibase 10.1038/ncomms2918} {\bibfield  {journal} {\bibinfo  {journal}
  {Nat. Comm.}\ }\textbf {\bibinfo {volume} {4}},\ \bibinfo {pages} {1887}
  (\bibinfo {year} {2013})}\BibitemShut {NoStop}%
\bibitem [{\citenamefont {Valeriani}, \citenamefont {Sanz},\ and\ \citenamefont
  {Frenkel}(2005)}]{valeriani_rate_2005}%
  \BibitemOpen
  \bibfield  {author} {\bibinfo {author} {\bibfnamefont {C.}~\bibnamefont
  {Valeriani}}, \bibinfo {author} {\bibfnamefont {E.}~\bibnamefont {Sanz}}, \
  and\ \bibinfo {author} {\bibfnamefont {D.}~\bibnamefont {Frenkel}},\ }\href
  {\doibase 10.1063/1.1896348} {\bibfield  {journal} {\bibinfo  {journal} {J.
  Chem. Phys.}\ }\textbf {\bibinfo {volume} {122}},\ \bibinfo {pages} {194501}
  (\bibinfo {year} {2005})}\BibitemShut {NoStop}%
\bibitem [{\citenamefont {Wang}, \citenamefont {Valeriani},\ and\ \citenamefont
  {Frenkel}(2009)}]{wang_homogeneous_2009}%
  \BibitemOpen
  \bibfield  {author} {\bibinfo {author} {\bibfnamefont {Z.-J.}\ \bibnamefont
  {Wang}}, \bibinfo {author} {\bibfnamefont {C.}~\bibnamefont {Valeriani}}, \
  and\ \bibinfo {author} {\bibfnamefont {D.}~\bibnamefont {Frenkel}},\ }\href
  {\doibase 10.1021/jp807727p} {\bibfield  {journal} {\bibinfo  {journal} {J.
  Phys. Chem. B}\ }\textbf {\bibinfo {volume} {113}},\ \bibinfo {pages} {3776}
  (\bibinfo {year} {2009})}\BibitemShut {NoStop}%
\bibitem [{\citenamefont {Bi}\ and\ \citenamefont
  {Li}(2014)}]{bi_probing_2014}%
  \BibitemOpen
  \bibfield  {author} {\bibinfo {author} {\bibfnamefont {Y.}~\bibnamefont
  {Bi}}\ and\ \bibinfo {author} {\bibfnamefont {T.}~\bibnamefont {Li}},\ }\href
  {\doibase 10.1021/jp503000u} {\bibfield  {journal} {\bibinfo  {journal} {J.
  Phys. Chem. B}\ }\textbf {\bibinfo {volume} {118}},\ \bibinfo {pages} {13324}
  (\bibinfo {year} {2014})}\BibitemShut {NoStop}%
\bibitem [{\citenamefont {Gianetti}\ \emph {et~al.}(2016)\citenamefont
  {Gianetti}, \citenamefont {Haji-Akbari}, \citenamefont {Longinotti},\ and\
  \citenamefont {Debenedetti}}]{gianetti_computational_2016}%
  \BibitemOpen
  \bibfield  {author} {\bibinfo {author} {\bibfnamefont {M.~M.}\ \bibnamefont
  {Gianetti}}, \bibinfo {author} {\bibfnamefont {A.}~\bibnamefont
  {Haji-Akbari}}, \bibinfo {author} {\bibfnamefont {M.~P.}\ \bibnamefont
  {Longinotti}}, \ and\ \bibinfo {author} {\bibfnamefont {P.~G.}\ \bibnamefont
  {Debenedetti}},\ }\href {\doibase 10.1039/C5CP06535F} {\bibfield  {journal}
  {\bibinfo  {journal} {Phys. Chem. Chem. Phys.}\ }\textbf {\bibinfo {volume}
  {18}},\ \bibinfo {pages} {4102} (\bibinfo {year} {2016})}\BibitemShut
  {NoStop}%
\bibitem [{\citenamefont {Moore}\ and\ \citenamefont
  {Molinero}(2011)}]{moore_is_2011}%
  \BibitemOpen
  \bibfield  {author} {\bibinfo {author} {\bibfnamefont {E.~B.}\ \bibnamefont
  {Moore}}\ and\ \bibinfo {author} {\bibfnamefont {V.}~\bibnamefont
  {Molinero}},\ }\href {\doibase 10.1039/c1cp22022e} {\bibfield  {journal}
  {\bibinfo  {journal} {Phys. Chem. Chem. Phys.}\ }\textbf {\bibinfo {volume}
  {13}},\ \bibinfo {pages} {20008} (\bibinfo {year} {2011})}\BibitemShut
  {NoStop}%
\bibitem [{\citenamefont {Hansen}\ \emph {et~al.}(2008)\citenamefont {Hansen},
  \citenamefont {Koza}, \citenamefont {Lindner},\ and\ \citenamefont
  {Kuhs}}]{hansen_formation_2008}%
  \BibitemOpen
  \bibfield  {author} {\bibinfo {author} {\bibfnamefont {T.~C.}\ \bibnamefont
  {Hansen}}, \bibinfo {author} {\bibfnamefont {M.~M.}\ \bibnamefont {Koza}},
  \bibinfo {author} {\bibfnamefont {P.}~\bibnamefont {Lindner}}, \ and\
  \bibinfo {author} {\bibfnamefont {W.~F.}\ \bibnamefont {Kuhs}},\ }\href
  {\doibase 10.1088/0953-8984/20/28/285105} {\bibfield  {journal} {\bibinfo
  {journal} {J. Phys. Cond. Matt.}\ }\textbf {\bibinfo {volume} {20}},\
  \bibinfo {pages} {285105} (\bibinfo {year} {2008})}\BibitemShut {NoStop}%
\bibitem [{\citenamefont {Malkin}\ \emph {et~al.}(2014)\citenamefont {Malkin},
  \citenamefont {Murray}, \citenamefont {Salzmann}, \citenamefont {Molinero},
  \citenamefont {Pickering},\ and\ \citenamefont
  {Whale}}]{malkin_stacking_2014}%
  \BibitemOpen
  \bibfield  {author} {\bibinfo {author} {\bibfnamefont {T.~L.}\ \bibnamefont
  {Malkin}}, \bibinfo {author} {\bibfnamefont {B.~J.}\ \bibnamefont {Murray}},
  \bibinfo {author} {\bibfnamefont {C.~G.}\ \bibnamefont {Salzmann}}, \bibinfo
  {author} {\bibfnamefont {V.}~\bibnamefont {Molinero}}, \bibinfo {author}
  {\bibfnamefont {S.~J.}\ \bibnamefont {Pickering}}, \ and\ \bibinfo {author}
  {\bibfnamefont {T.~F.}\ \bibnamefont {Whale}},\ }\href {\doibase
  10.1039/C4CP02893G} {\bibfield  {journal} {\bibinfo  {journal} {Phys. Chem.
  Chem. Phys.}\ }\textbf {\bibinfo {volume} {17}},\ \bibinfo {pages} {60}
  (\bibinfo {year} {2014})}\BibitemShut {NoStop}%
\bibitem [{\citenamefont {Sear}(2007)}]{sear_nucleation:_2007}%
  \BibitemOpen
  \bibfield  {author} {\bibinfo {author} {\bibfnamefont {R.~P.}\ \bibnamefont
  {Sear}},\ }\href {\doibase 10.1088/0953-8984/19/3/033101} {\bibfield
  {journal} {\bibinfo  {journal} {J. Phys. Cond. Matt.}\ }\textbf {\bibinfo
  {volume} {19}},\ \bibinfo {pages} {033101} (\bibinfo {year}
  {2007})}\BibitemShut {NoStop}%
\bibitem [{\citenamefont {Auer}\ and\ \citenamefont
  {Frenkel}(2003)}]{auer_line_2003}%
  \BibitemOpen
  \bibfield  {author} {\bibinfo {author} {\bibfnamefont {S.}~\bibnamefont
  {Auer}}\ and\ \bibinfo {author} {\bibfnamefont {D.}~\bibnamefont {Frenkel}},\
  }\href {\doibase 10.1103/PhysRevLett.91.015703} {\bibfield  {journal}
  {\bibinfo  {journal} {Phys. Rev. Lett.}\ }\textbf {\bibinfo {volume} {91}},\
  \bibinfo {pages} {015703} (\bibinfo {year} {2003})}\BibitemShut {NoStop}%
\bibitem [{\citenamefont {Winter}, \citenamefont {Virnau},\ and\ \citenamefont
  {Binder}(2009)}]{winter_monte_2009}%
  \BibitemOpen
  \bibfield  {author} {\bibinfo {author} {\bibfnamefont {D.}~\bibnamefont
  {Winter}}, \bibinfo {author} {\bibfnamefont {P.}~\bibnamefont {Virnau}}, \
  and\ \bibinfo {author} {\bibfnamefont {K.}~\bibnamefont {Binder}},\ }\href
  {\doibase 10.1103/PhysRevLett.103.225703} {\bibfield  {journal} {\bibinfo
  {journal} {Phys. Rev. Lett.}\ }\textbf {\bibinfo {volume} {103}},\ \bibinfo
  {pages} {225703} (\bibinfo {year} {2009})}\BibitemShut {NoStop}%
\bibitem [{\citenamefont {Sellberg}\ \emph {et~al.}(2014)\citenamefont
  {Sellberg}, \citenamefont {Huang}, \citenamefont {McQueen}, \citenamefont
  {Loh}, \citenamefont {Laksmono}, \citenamefont {Schlesinger}, \citenamefont
  {Sierra}, \citenamefont {Nordlund}, \citenamefont {Hampton}, \citenamefont
  {Starodub}, \citenamefont {DePonte}, \citenamefont {Beye}, \citenamefont
  {Chen}, \citenamefont {Martin}, \citenamefont {Barty}, \citenamefont
  {Wikfeldt}, \citenamefont {Weiss}, \citenamefont {Caronna}, \citenamefont
  {Feldkamp}, \citenamefont {Skinner}, \citenamefont {Seibert}, \citenamefont
  {Messerschmidt}, \citenamefont {Williams}, \citenamefont {Boutet},
  \citenamefont {Pettersson}, \citenamefont {Bogan},\ and\ \citenamefont
  {Nilsson}}]{sellberg_ultrafast_2014}%
  \BibitemOpen
  \bibfield  {author} {\bibinfo {author} {\bibfnamefont {J.~A.}\ \bibnamefont
  {Sellberg}}, \bibinfo {author} {\bibfnamefont {C.}~\bibnamefont {Huang}},
  \bibinfo {author} {\bibfnamefont {T.~A.}\ \bibnamefont {McQueen}}, \bibinfo
  {author} {\bibfnamefont {N.~D.}\ \bibnamefont {Loh}}, \bibinfo {author}
  {\bibfnamefont {H.}~\bibnamefont {Laksmono}}, \bibinfo {author}
  {\bibfnamefont {D.}~\bibnamefont {Schlesinger}}, \bibinfo {author}
  {\bibfnamefont {R.~G.}\ \bibnamefont {Sierra}}, \bibinfo {author}
  {\bibfnamefont {D.}~\bibnamefont {Nordlund}}, \bibinfo {author}
  {\bibfnamefont {C.~Y.}\ \bibnamefont {Hampton}}, \bibinfo {author}
  {\bibfnamefont {D.}~\bibnamefont {Starodub}}, \bibinfo {author}
  {\bibfnamefont {D.~P.}\ \bibnamefont {DePonte}}, \bibinfo {author}
  {\bibfnamefont {M.}~\bibnamefont {Beye}}, \bibinfo {author} {\bibfnamefont
  {C.}~\bibnamefont {Chen}}, \bibinfo {author} {\bibfnamefont {A.~V.}\
  \bibnamefont {Martin}}, \bibinfo {author} {\bibfnamefont {A.}~\bibnamefont
  {Barty}}, \bibinfo {author} {\bibfnamefont {K.~T.}\ \bibnamefont {Wikfeldt}},
  \bibinfo {author} {\bibfnamefont {T.~M.}\ \bibnamefont {Weiss}}, \bibinfo
  {author} {\bibfnamefont {C.}~\bibnamefont {Caronna}}, \bibinfo {author}
  {\bibfnamefont {J.}~\bibnamefont {Feldkamp}}, \bibinfo {author}
  {\bibfnamefont {L.~B.}\ \bibnamefont {Skinner}}, \bibinfo {author}
  {\bibfnamefont {M.~M.}\ \bibnamefont {Seibert}}, \bibinfo {author}
  {\bibfnamefont {M.}~\bibnamefont {Messerschmidt}}, \bibinfo {author}
  {\bibfnamefont {G.~J.}\ \bibnamefont {Williams}}, \bibinfo {author}
  {\bibfnamefont {S.}~\bibnamefont {Boutet}}, \bibinfo {author} {\bibfnamefont
  {L.~G.~M.}\ \bibnamefont {Pettersson}}, \bibinfo {author} {\bibfnamefont
  {M.~J.}\ \bibnamefont {Bogan}}, \ and\ \bibinfo {author} {\bibfnamefont
  {A.}~\bibnamefont {Nilsson}},\ }\href {\doibase 10.1038/nature13266}
  {\bibfield  {journal} {\bibinfo  {journal} {Nature}\ }\textbf {\bibinfo
  {volume} {510}},\ \bibinfo {pages} {381} (\bibinfo {year}
  {2014})}\BibitemShut {NoStop}%
\bibitem [{\citenamefont {Sosso}\ \emph {et~al.}(2016)\citenamefont {Sosso},
  \citenamefont {Chen}, \citenamefont {Cox}, \citenamefont {Fitzner},
  \citenamefont {Pedevilla}, \citenamefont {Zen},\ and\ \citenamefont
  {Michaelides}}]{sosso_cnm}%
  \BibitemOpen
  \bibfield  {author} {\bibinfo {author} {\bibfnamefont {G.}~\bibnamefont
  {Sosso}}, \bibinfo {author} {\bibfnamefont {J.}~\bibnamefont {Chen}},
  \bibinfo {author} {\bibfnamefont {S.}~\bibnamefont {Cox}}, \bibinfo {author}
  {\bibfnamefont {M.}~\bibnamefont {Fitzner}}, \bibinfo {author} {\bibfnamefont
  {P.}~\bibnamefont {Pedevilla}}, \bibinfo {author} {\bibfnamefont
  {A.}~\bibnamefont {Zen}}, \ and\ \bibinfo {author} {\bibfnamefont
  {A.}~\bibnamefont {Michaelides}},\ }\href@noop {} {\bibfield  {journal}
  {\bibinfo  {journal} {Chem. Rev.}\ ,\ \bibinfo {pages} {DOI:
  10.1021/acs.chemrev.5b00744}} (\bibinfo {year} {2016})}\BibitemShut {NoStop}%
\bibitem [{\citenamefont {Ickes}, \citenamefont {Welti},\ and\ \citenamefont
  {Lohmann}(2016)}]{ickes_classical_2016}%
  \BibitemOpen
  \bibfield  {author} {\bibinfo {author} {\bibfnamefont {L.}~\bibnamefont
  {Ickes}}, \bibinfo {author} {\bibfnamefont {A.}~\bibnamefont {Welti}}, \ and\
  \bibinfo {author} {\bibfnamefont {U.}~\bibnamefont {Lohmann}},\ }\href
  {\doibase 10.5194/acp-2015-969} {\bibfield  {journal} {\bibinfo  {journal}
  {Atm. Chem. Phys. Discuss.}\ }\textbf {\bibinfo {volume} {16}},\ \bibinfo
  {pages} {1} (\bibinfo {year} {2016})}\BibitemShut {NoStop}%
\bibitem [{\citenamefont {Ehre}\ \emph {et~al.}(2010)\citenamefont {Ehre},
  \citenamefont {Lavert}, \citenamefont {Lahav},\ and\ \citenamefont
  {Lubomirsky}}]{ehre_water_2010}%
  \BibitemOpen
  \bibfield  {author} {\bibinfo {author} {\bibfnamefont {D.}~\bibnamefont
  {Ehre}}, \bibinfo {author} {\bibfnamefont {E.}~\bibnamefont {Lavert}},
  \bibinfo {author} {\bibfnamefont {M.}~\bibnamefont {Lahav}}, \ and\ \bibinfo
  {author} {\bibfnamefont {I.}~\bibnamefont {Lubomirsky}},\ }\href {\doibase
  10.1126/science.1178085} {\bibfield  {journal} {\bibinfo  {journal}
  {Science}\ }\textbf {\bibinfo {volume} {327}},\ \bibinfo {pages} {672}
  (\bibinfo {year} {2010})}\BibitemShut {NoStop}%
\bibitem [{\citenamefont {Yan}\ and\ \citenamefont
  {Patey}(2011)}]{yan_heterogeneous_2011}%
  \BibitemOpen
  \bibfield  {author} {\bibinfo {author} {\bibfnamefont {J.~Y.}\ \bibnamefont
  {Yan}}\ and\ \bibinfo {author} {\bibfnamefont {G.~N.}\ \bibnamefont
  {Patey}},\ }\href {\doibase 10.1021/jz201113m} {\bibfield  {journal}
  {\bibinfo  {journal} {J. Phys. Chem. Letters}\ }\textbf {\bibinfo {volume}
  {2}},\ \bibinfo {pages} {2555} (\bibinfo {year} {2011})}\BibitemShut
  {NoStop}%
\bibitem [{\citenamefont {Carrasco}, \citenamefont {Hodgson},\ and\
  \citenamefont {Michaelides}(2012)}]{carrasco_molecular_2012}%
  \BibitemOpen
  \bibfield  {author} {\bibinfo {author} {\bibfnamefont {J.}~\bibnamefont
  {Carrasco}}, \bibinfo {author} {\bibfnamefont {A.}~\bibnamefont {Hodgson}}, \
  and\ \bibinfo {author} {\bibfnamefont {A.}~\bibnamefont {Michaelides}},\
  }\href {\doibase 10.1038/nmat3354} {\bibfield  {journal} {\bibinfo  {journal}
  {Nat. Mat.}\ }\textbf {\bibinfo {volume} {11}},\ \bibinfo {pages} {667}
  (\bibinfo {year} {2012})}\BibitemShut {NoStop}%
\end{thebibliography}

\begin{mcitethebibliography}{24}
\providecommand*\natexlab[1]{#1}
\providecommand*\mciteSetBstSublistMode[1]{}
\providecommand*\mciteSetBstMaxWidthForm[2]{}
\providecommand*\mciteBstWouldAddEndPuncttrue
  {\def\EndOfBibitem{\unskip.}}
\providecommand*\mciteBstWouldAddEndPunctfalse
  {\let\EndOfBibitem\relax}
\providecommand*\mciteSetBstMidEndSepPunct[3]{}
\providecommand*\mciteSetBstSublistLabelBeginEnd[3]{}
\providecommand*\EndOfBibitem{}
\mciteSetBstSublistMode{f}
\mciteSetBstMaxWidthForm{subitem}{(\alph{mcitesubitemcount})}
\mciteSetBstSublistLabelBeginEnd
  {\mcitemaxwidthsubitemform\space}
  {\relax}
  {\relax}

\bibitem[Bish(1993)]{bish_rietveld_1993}
Bish,~D.~L. \emph{Clays and Clay Minerals} \textbf{1993}, \emph{41},
  738--744\relax
\mciteBstWouldAddEndPuncttrue
\mciteSetBstMidEndSepPunct{\mcitedefaultmidpunct}
{\mcitedefaultendpunct}{\mcitedefaultseppunct}\relax
\EndOfBibitem
\bibitem[Abascal \latin{et~al.}(2005)Abascal, Sanz, Fernandez, and
  Vega]{abascal_potential_2005}
Abascal,~J. L.~F.; Sanz,~E.; Fernandez,~R.~G.; Vega,~C. \emph{The Journal of
  Chemical Physics} \textbf{2005}, \emph{122}, 234511\relax
\mciteBstWouldAddEndPuncttrue
\mciteSetBstMidEndSepPunct{\mcitedefaultmidpunct}
{\mcitedefaultendpunct}{\mcitedefaultseppunct}\relax
\EndOfBibitem
\bibitem[Zielke \latin{et~al.}(2016)Zielke, Bertram, and
  Patey]{zielke_simulations_2016}
Zielke,~S.~A.; Bertram,~A.~K.; Patey,~G.~N. \emph{J. Phys. Chem. B}
  \textbf{2016}, \emph{120}, 1726--1734\relax
\mciteBstWouldAddEndPuncttrue
\mciteSetBstMidEndSepPunct{\mcitedefaultmidpunct}
{\mcitedefaultendpunct}{\mcitedefaultseppunct}\relax
\EndOfBibitem
\bibitem[Cygan \latin{et~al.}(2004)Cygan, Liang, and
  Kalinichev]{cygan_molecular_2004}
Cygan,~R.~T.; Liang,~J.-J.; Kalinichev,~A.~G. \emph{The Journal of Physical
  Chemistry B} \textbf{2004}, \emph{108}, 1255--1266\relax
\mciteBstWouldAddEndPuncttrue
\mciteSetBstMidEndSepPunct{\mcitedefaultmidpunct}
{\mcitedefaultendpunct}{\mcitedefaultseppunct}\relax
\EndOfBibitem
\bibitem[Haji-Akbari and Debenedetti(2015)Haji-Akbari, and
  Debenedetti]{haji-akbari_direct_2015}
Haji-Akbari,~A.; Debenedetti,~P.~G. \emph{Proceedings of the National Academy
  of Sciences} \textbf{2015}, 201509267\relax
\mciteBstWouldAddEndPuncttrue
\mciteSetBstMidEndSepPunct{\mcitedefaultmidpunct}
{\mcitedefaultendpunct}{\mcitedefaultseppunct}\relax
\EndOfBibitem
\bibitem[Lorentz(1881)]{lorentz_ueber_1881}
Lorentz,~H.~A. \emph{Annalen der Physik} \textbf{1881}, \emph{248},
  127--136\relax
\mciteBstWouldAddEndPuncttrue
\mciteSetBstMidEndSepPunct{\mcitedefaultmidpunct}
{\mcitedefaultendpunct}{\mcitedefaultseppunct}\relax
\EndOfBibitem
\bibitem[Berthelot and Hebd(1898)Berthelot, and Hebd]{ahthefrench}
Berthelot,~D.; Hebd,~C.~R. Seances De L' Academie Des Sciences. 1898\relax
\mciteBstWouldAddEndPuncttrue
\mciteSetBstMidEndSepPunct{\mcitedefaultmidpunct}
{\mcitedefaultendpunct}{\mcitedefaultseppunct}\relax
\EndOfBibitem
\bibitem[Bussi \latin{et~al.}(2007)Bussi, Donadio, and
  Parrinello]{bussi_canonical_2007}
Bussi,~G.; Donadio,~D.; Parrinello,~M. \emph{The Journal of Chemical Physics}
  \textbf{2007}, \emph{126}, 014101\relax
\mciteBstWouldAddEndPuncttrue
\mciteSetBstMidEndSepPunct{\mcitedefaultmidpunct}
{\mcitedefaultendpunct}{\mcitedefaultseppunct}\relax
\EndOfBibitem
\bibitem[Miyamoto and Kollman(1992)Miyamoto, and
  Kollman]{miyamoto_settle:_1992}
Miyamoto,~S.; Kollman,~P.~A. \emph{Journal of Computational Chemistry}
  \textbf{1992}, \emph{13}, 952--962\relax
\mciteBstWouldAddEndPuncttrue
\mciteSetBstMidEndSepPunct{\mcitedefaultmidpunct}
{\mcitedefaultendpunct}{\mcitedefaultseppunct}\relax
\EndOfBibitem
\bibitem[Hess \latin{et~al.}(1997)Hess, Bekker, Berendsen, and
  Fraaije]{hess_lincs:_1997}
Hess,~B.; Bekker,~H.; Berendsen,~H. J.~C.; Fraaije,~J. G. E.~M. \emph{Journal
  of Computational Chemistry} \textbf{1997}, \emph{18}, 1463--1472\relax
\mciteBstWouldAddEndPuncttrue
\mciteSetBstMidEndSepPunct{\mcitedefaultmidpunct}
{\mcitedefaultendpunct}{\mcitedefaultseppunct}\relax
\EndOfBibitem
\bibitem[Steinhardt \latin{et~al.}(1983)Steinhardt, Nelson, and
  Ronchetti]{steinhardt_bond-orientational_1983}
Steinhardt,~P.~J.; Nelson,~D.~R.; Ronchetti,~M. \emph{Physical Review B}
  \textbf{1983}, \emph{28}, 784--805\relax
\mciteBstWouldAddEndPuncttrue
\mciteSetBstMidEndSepPunct{\mcitedefaultmidpunct}
{\mcitedefaultendpunct}{\mcitedefaultseppunct}\relax
\EndOfBibitem
\bibitem[Tribello \latin{et~al.}(2014)Tribello, Bonomi, Branduardi, Camilloni,
  and Bussi]{tribello_plumed_2014}
Tribello,~G.~A.; Bonomi,~M.; Branduardi,~D.; Camilloni,~C.; Bussi,~G.
  \emph{Computer Physics Communications} \textbf{2014}, \emph{185},
  604--613\relax
\mciteBstWouldAddEndPuncttrue
\mciteSetBstMidEndSepPunct{\mcitedefaultmidpunct}
{\mcitedefaultendpunct}{\mcitedefaultseppunct}\relax
\EndOfBibitem
\bibitem[Bi and Li(2014)Bi, and Li]{bi_probing_2014}
Bi,~Y.; Li,~T. \emph{The Journal of Physical Chemistry B} \textbf{2014},
  \emph{118}, 13324--13332\relax
\mciteBstWouldAddEndPuncttrue
\mciteSetBstMidEndSepPunct{\mcitedefaultmidpunct}
{\mcitedefaultendpunct}{\mcitedefaultseppunct}\relax
\EndOfBibitem
\bibitem[Allen \latin{et~al.}(2006)Allen, Frenkel, and
  Wolde]{allen_forward_2006}
Allen,~R.~J.; Frenkel,~D.; Wolde,~P. R.~t. \emph{The Journal of Chemical
  Physics} \textbf{2006}, \emph{124}, 194111\relax
\mciteBstWouldAddEndPuncttrue
\mciteSetBstMidEndSepPunct{\mcitedefaultmidpunct}
{\mcitedefaultendpunct}{\mcitedefaultseppunct}\relax
\EndOfBibitem
\bibitem[Sear(2007)]{sear_nucleation:_2007}
Sear,~R.~P. \emph{Journal of Physics: Condensed Matter} \textbf{2007},
  \emph{19}, 033101\relax
\mciteBstWouldAddEndPuncttrue
\mciteSetBstMidEndSepPunct{\mcitedefaultmidpunct}
{\mcitedefaultendpunct}{\mcitedefaultseppunct}\relax
\EndOfBibitem
\bibitem[Kalikmanov(2013)]{kalikmanov_nucleation_2013}
Kalikmanov,~V. \emph{Nucleation {Theory}}; Lecture {Notes} in {Physics};
  Springer Netherlands: Dordrecht, 2013; Vol. 860\relax
\mciteBstWouldAddEndPuncttrue
\mciteSetBstMidEndSepPunct{\mcitedefaultmidpunct}
{\mcitedefaultendpunct}{\mcitedefaultseppunct}\relax
\EndOfBibitem
\bibitem[Li \latin{et~al.}(2011)Li, Donadio, Russo, and
  Galli]{li_homogeneous_2011}
Li,~T.; Donadio,~D.; Russo,~G.; Galli,~G. \emph{Physical Chemistry Chemical
  Physics} \textbf{2011}, \emph{13}, 19807--19813\relax
\mciteBstWouldAddEndPuncttrue
\mciteSetBstMidEndSepPunct{\mcitedefaultmidpunct}
{\mcitedefaultendpunct}{\mcitedefaultseppunct}\relax
\EndOfBibitem
\bibitem[Li \latin{et~al.}(2013)Li, Donadio, and Galli]{li_ice_2013}
Li,~T.; Donadio,~D.; Galli,~G. \emph{Nature Communications} \textbf{2013},
  \emph{4}, 1887\relax
\mciteBstWouldAddEndPuncttrue
\mciteSetBstMidEndSepPunct{\mcitedefaultmidpunct}
{\mcitedefaultendpunct}{\mcitedefaultseppunct}\relax
\EndOfBibitem
\bibitem[Gianetti \latin{et~al.}(2016)Gianetti, Haji-Akbari, Longinotti, and
  Debenedetti]{gianetti_computational_2016}
Gianetti,~M.~M.; Haji-Akbari,~A.; Longinotti,~M.~P.; Debenedetti,~P.~G.
  \emph{Physical Chemistry Chemical Physics} \textbf{2016}, \emph{18},
  4102--4111\relax
\mciteBstWouldAddEndPuncttrue
\mciteSetBstMidEndSepPunct{\mcitedefaultmidpunct}
{\mcitedefaultendpunct}{\mcitedefaultseppunct}\relax
\EndOfBibitem
\bibitem[King(1967)]{king_ring_1967}
King,~S.~V. \emph{Nature} \textbf{1967}, \emph{213}, 1112--1113\relax
\mciteBstWouldAddEndPuncttrue
\mciteSetBstMidEndSepPunct{\mcitedefaultmidpunct}
{\mcitedefaultendpunct}{\mcitedefaultseppunct}\relax
\EndOfBibitem
\bibitem[Franzblau(1991)]{franzblau_computation_1991}
Franzblau,~D.~S. \emph{Physical Review B} \textbf{1991}, \emph{44},
  4925--4930\relax
\mciteBstWouldAddEndPuncttrue
\mciteSetBstMidEndSepPunct{\mcitedefaultmidpunct}
{\mcitedefaultendpunct}{\mcitedefaultseppunct}\relax
\EndOfBibitem
\bibitem[Le~Roux and Jund(2010)Le~Roux, and Jund]{le_roux_ring_2010}
Le~Roux,~S.; Jund,~P. \emph{Computational Materials Science} \textbf{2010},
  \emph{49}, 70--83\relax
\mciteBstWouldAddEndPuncttrue
\mciteSetBstMidEndSepPunct{\mcitedefaultmidpunct}
{\mcitedefaultendpunct}{\mcitedefaultseppunct}\relax
\EndOfBibitem
\bibitem[Rawat and Biswas(2011)Rawat, and Biswas]{rawat_shape_2011}
Rawat,~N.; Biswas,~P. \emph{Physical Chemistry Chemical Physics} \textbf{2011},
  \emph{13}, 9632--9643\relax
\mciteBstWouldAddEndPuncttrue
\mciteSetBstMidEndSepPunct{\mcitedefaultmidpunct}
{\mcitedefaultendpunct}{\mcitedefaultseppunct}\relax
\EndOfBibitem
\end{mcitethebibliography}
\end{document}